\documentclass[12pt]{article}
\pdfoutput=1

\usepackage[a4paper]{geometry}
\usepackage[T1]{fontenc} 
\usepackage{comment}
\usepackage{jheppub}
\usepackage{hyperref,float,array,adjustbox,mathtools,physics,xcolor}
\usepackage{csquotes}
\usepackage{lipsum}
\usepackage{booktabs}
\usepackage{pdflscape}
\usepackage{geometry}
\usepackage{fancyhdr}
\usepackage{changepage}
\usepackage[english]{babel}
\usepackage{xcolor,tikz,pgfplots,amsmath,amsthm,amsfonts}
\usepackage{subcaption}
\usetikzlibrary{arrows.meta} % punte freccia moderne
\pgfplotsset{compat=1.18} % o compat=newest se ce l'hai
\usepackage{graphicx}
\usepackage[mathscr]{euscript}
\usetikzlibrary{matrix,calc,positioning,decorations.markings,decorations.pathmorphing,decorations.pathreplacing}
\usetikzlibrary{arrows,cd} 
\usetikzlibrary{shapes.misc}
\usetikzlibrary {shapes.geometric}
\usetikzlibrary {shapes.multipart}
\usetikzlibrary{decorations.markings}
\usetikzlibrary{backgrounds}
\usetikzlibrary{shapes.gates.logic.US}
\usetikzlibrary{circuits.ee.IEC}
\usepackage{bbding}
\usepackage{tikz}
\usetikzlibrary{automata,positioning,calc}
\usetikzlibrary{decorations.markings}
\tikzset{->-/.style={decoration={markings, mark=at position #1 with {\arrow{>}}},postaction={decorate}}}
\tikzset{-<-/.style={decoration={markings, mark=at position #1 with {\arrow{<}}},postaction={decorate}}}
\tikzset{auto shift/.style={auto=right,->, to path={ let \p1=(\tikztostart),\p2=(\tikztotarget), \n1={atan2(\y2-\y1,\x2-\x1)},\n2={\n1+180} in ($(\tikztostart.{\n1})!1mm!270:(\tikztotarget.{\n2})$) -- ($(\tikztotarget.{\n2})!1mm!90:(\tikztostart.{\n1})$) \tikztonodes}}}

\DeclareMathOperator{\SU}{SU}
\DeclareMathOperator{\UU}{U}
\newcommand{\numberset}{\mathbb}

\newcommand{\Z}{\numberset{Z}}

\newcommand{\C}{\numberset{C}}
\newcommand{\T}{\numberset{T}}

\newcommand{\ee}{\mathrm{e}}
\newcommand{\mi}{\mathrm{i}}

\renewcommand{\Im}{\operatorname{Im}}

\newtheorem{conjecture}{Conjecture}
\definecolor{SUcol}{rgb}{0.5, 0.85, 0.8} 
\definecolor{USPcol}{rgb}{0.91,0.67,0.75}
\definecolor{SOcol}{rgb}{0.93,0.83,0.57}
\definecolor{Petrolio}{RGB}{60,99,113}
\definecolor{Prugna}{RGB}{150,65,95}
\definecolor{Senape}{RGB}{218, 165, 32}
\definecolor{Ardesia}{RGB}{96,125,139}
\definecolor{Lavanda}{RGB}{235,225,245}
\definecolor{Salvia}{RGB}{110,150,120}
\definecolor{Mattone}{RGB}{185,105,75}
\definecolor{MIXcol}{rgb}{0.705, 0.760, 0.775}
\definecolor{verde}{RGB}{80,190,90}

\title{
\begin{center}
New Bethe vacua for $\mathcal{N}=2$ elliptic models
\end{center}
}

\author[a]{Antonio Amariti,}	
\author[a,b]{Pietro Glorioso,}	
\author[a,b]{Chiara Mascherpa,}
\author[c]{and Andrea Zanetti}
\affiliation[a]{
	INFN, Sezione di Milano, Via Celoria 16, I-20133 Milano, Italy
	}
\affiliation[b]{
	Dipartimento di Fisica, Università degli studi di Milano, Via Celoria 16, I-20133
	}
\affiliation[c]{
	Physique Théorique et Mathématique and International Solvay Institutes
Université Libre de Bruxelles, C.P. 231, 1050 Brussels, Belgium
	}
\emailAdd{antonio.amariti@mi.infn.it}
\emailAdd{pietro.glorioso@mi.infn.it}
\emailAdd{chiara.mascherpa@mi.infn.it}
\emailAdd{andrea.zanetti@mi.infn.it}

\abstract{
We study the superconformal index of four-dimensional $\mathcal{N}=2$ $\mathrm{SU}(N)$ elliptic models through the Bethe Ansatz approach, where the index is evaluated as a sum of vacua arising from a set of transcendental equations.
Starting from the known discrete solutions of these equations, that we refer to as progenitors, we 
show that there exist large classes of new descendant solutions. 
We then show that the solutions can be organized into orbits of the S-duality group, 
mimicking its action  on the massive vacua of the $\mathcal{N}=1^*$ deformations of the elliptic spin Calogero-Moser models.
In this way we generalize the results already obtained for $\mathcal{N}=4$ $\mathrm{SU}(N)$ SYM, where a  relation between the  discrete solutions to the Bethe Ansatz Equations and the massive vacua of its $\mathcal{N}=1^*$ massive deformation was conjectured.  We conclude our analysis by determining the contributions of the new descendant solutions to the index.
}

\begin{document}
\maketitle
\flushbottom
\allowdisplaybreaks 
%
%
%
%
%%%%%%%%%%%%%%
%%%%%%%%%%%%%%
\section{Introduction}
\label{sec:intro}
%%%%%%%%%%%%%%
%%%%%%%%%%%%%%
%
%

The superconformal index (SCI) \cite{Kinney:2005ej,Romelsberger:2005eg} is a very useful quantity in the study of four dimensional supersymmetric gauge theories and it has been widely studied to extract exact non-perturbative information of the spectrum.

In recent years, the computation of this index has been formulated in terms of a Bethe problem \cite{Closset:2017bse,Benini:2018mlo} that consists of solving a set of transcendental algebraic equations, called Bethe Ansatz Equations (BAEs), and evaluating the SCI on these Bethe vacua.
Such a technique has been successfully applied in the large $N$ regime, where the Bethe vacua responsible for the leading contributions, called the Hong-Liu solutions (HL) \cite{Hosseini:2016cyf,Hong:2018viz}, are known.
This was the case for $\mathcal{N}=4$ $\SU(N)$ SYM theory in \cite{Benini:2018ywd, Aharony:2021zkr}, where the Bethe Ansatz approach was used to compute the SCI, providing a field-theoretic interpretation of the entropy of the holographic dual AdS black holes
\cite{Gutowski:2004yv,Gutowski:2004ez}.

For general values of $N$, the complete set of solutions to the BAEs of $\mathcal{N} = 4$ SYM is still unknown. Known solutions can be classified into two classes: discrete and continuous solutions.
\begin{itemize}
\item
The discrete solutions correspond to isolated points $\hat{u}$ in the space of holonomies. A large known family of discrete solutions is provided by the HL  solutions. In the large $N$ limit, each solution is holographically associated with a distinct Euclidean gravitational saddle. In this regime, the index is expected to reproduce the microstate counting of the $\frac{1}{16}$ BPS charged rotating $\mathrm{AdS}_5$ black hole, first introduced by \cite{Gutowski:2004yv}
(see \cite{Zaffaroni:2019dhb} for review).
More precisely in \cite{Hosseini:2017mds}
an entropy function that accounts for the dual
black hole microstates was originally proposed. Then the microscopic origin of the  entropy
of the  dual  black hole was provided in \cite{Cabo-Bizet:2018ehj}.
In \cite{Benini:2018ywd} the SCI was evaluated in the large $N$ limit via the BA formula \eqref{baf}, and the dominant contribution was found to arise from a solution known as the basic solution.
The logarithm of its contribution reproduces the entropy function of the dual black hole geometry \cite{Hosseini:2017mds}, extending the result obtained in the Cardy-like limit\footnote{See also \cite{Honda:2019cio, ArabiArdehali:2019tdm,Kim:2019yrz,Cabo-Bizet:2019osg,Amariti:2019mgp,Cabo-Bizet:2019eaf,ArabiArdehali:2019orz,Murthy:2020scj,Cabo-Bizet:2020nkr,Agarwal:2020zwm,Benini:2020gjh,GonzalezLezcano:2020yeb,Copetti:2020dil,Amariti:2020jyx,Amariti:2021ubd,Cassani:2021fyv,ArabiArdehali:2021nsx,Jejjala:2021hlt,Colombo:2021kbb,Cabo-Bizet:2021plf,Mamroud:2022msu,Amariti:2023rci,Choi:2023tiq,Aharony:2024ntg} for other results about the relation between the 4d SCI and the entropy of the dual black holes.} at small $
\tau$ \cite{Choi:2018hmj} to arbitrary $
\tau$. 
Furthermore, in the case of $\SU(N)$ $\mathcal{N}=4$ SYM,
the correspondence between HL solutions and black hole geometries was further extended in \cite{Aharony:2021zkr}, where the logarithm of each HL contribution to the BA formula was shown to reproduce the corresponding dual entropy function.
\item
Continuous solutions, on the other hand, correspond to complex manifolds in the space of holonomies. In \cite{Cabo-Bizet:2024kfe}, a generalization of the BA formula \eqref{baf} to continuous solutions was proposed, showing that their contributions localize on a finite set of points\footnote{
See also \cite{Closset:2017bse,Fazzi:2026wkb} for further comments on continuous solutions.}.
As shown in \cite{Benini:2021ano, Lezcano:2021qbj} the HL solutions fully reproduce the SCI in the case of $\SU(2)$. 
However, for $N \ge 3$, the contribution of continuous solutions is crucial to retrieve the full SCI. In the case of $\SU(3)$ the sum of the contributions of the discrete solutions and those of a family of continuous solutions fully reproduces the SCI \cite{Cabo-Bizet:2024kfe}. Beyond $N = 3$ numerical evidence to $N = 10$ suggests that there are $\ell$-complex dimensional branches of continuous solutions for $N\geq(\ell+1)(\ell+2)/2$.
\end{itemize}

The HL solutions are the only known discrete solutions  with non-vanishing contributions to the SCI so far. However, there is no proof of the fact that they are the only discrete solutions that contribute to the SCI. Moreover, as discussed above, the explicit form of the continuous solutions is known only for $N=3$. 
The exact evaluation of the SCI at finite rank remains then an open question, because the contribution of solutions beyond the HL family is required to reproduce the full index \cite{Benini:2021ano,Lezcano:2021qbj,Cabo-Bizet:2024kfe,Amariti:2025vjd,Fazzi:2026fao}.
These additional solutions can either be discrete, as in the case of the HL solutions, or continuous. 

Restricting to the HL solutions 
of the $\mathrm{SU}(N)$ $\mathcal{N} = 4$ SYM case,
they  can be put in 1-1 correspondence with 
the  vacua of the massive $\mathcal{N}=1^*$  deformations of $\mathrm{SU}(N)$  $\mathcal{N}=4$ SYM.
A conjecture relating these two apparently unrelated classifications 
was originally proposed in \cite{ArabiArdehali:2019orz} and later refined in \cite{Benini:2021ano} (see also \cite{Fazzi:2026wkb} for recent developments beyond the $\SU(N)$ case).  Explicitly the conjecture  relates the solutions of the BAEs of $\SU(N)$ $\mathcal{N}=4$ SYM to the vacua of the elliptic Calogero-Moser integrable system arising from a massive deformation of the $\mathcal{N}=4$ SYM theory \cite{Donagi:1995cf,Dorey:1999sj}
(see also \cite{Bourget:2015cza,Bourget:2015lua,Bourget:2015upj,Bourget:2016yhy,Bourget:2017goy,Damia:2025bla} for recent developments).
More precisely, the conjecture of \cite{Benini:2021ano} can be stated as follows:
\begin{conjecture}
\label{Conjecture}
    Solutions to the Bethe Ansatz Equations of $\mathrm{SU}(N)$ $\mathcal{N} = 4$ SYM which provide non-vanishing contribution to the superconformal index are in one-to-one correspondence with the vacua of the $\mathcal{N}=1^*$ massive deformation of the theory on $\mathbb{R}^{1,3}$:
    \begin{equation*}
	\begin{aligned}
		\text{discrete solutions} \quad
		& \longleftrightarrow \quad
		\text{massive vacua}\,, \\
		\text{ $\mathbb
        {C}^\ell$-dim continua of vacua} \quad
		& \longleftrightarrow \quad
		\text{branches of Coulomb vacua with $\ell$ photons}\,.
	\end{aligned}
\end{equation*}
\end{conjecture}
Then, in this correspondence, discrete solutions correspond to isolated massive vacua, while continuous families of solutions correspond to vacua with flat directions. 
Focusing on the relation between discrete solutions and massive vacua, one can see that the modularity properties of the BAEs reproduce the action of the inherited S-duality group between the massive vacua. 
In other words, starting from the \emph{basic} HL solution \cite{Hosseini:2016cyf,Hong:2018viz} and acting on the rotational parameter $\tau(=\sigma$) and on the gauge holonomies with two generators, $S$ and $T$, one reproduces all the HL solutions.
The inequivalent solutions correspond to the massive vacua of the integrable system and an S-duality orbit is generated by the action of the inherited \cite{Leigh:1995ep,Argyres:1999xu}  modular group at the level of the Calogero-Moser system.

One can wonder if 
the conjecture can be extended 
beyond the case of maximal supersymmetry.
For example  one can look for 4d models with  eight supercharges admitting   Calogero-Moser type integrable systems, once  proper $\mathcal{N}=1^*$ massive deformations are turned on.
Such models are 
$\mathcal{N}=2$ necklace quivers with $\SU(N)$ gauge groups,
obtained by compactification of the 6d $\mathcal{N}=(2,0)$ theory on a punctured torus \cite{Donagi:1995cf}. 
The relation with the Calogero-Moser systems and the counting of massive vacua in these cases were performed in \cite{Dorey:2001qj,Hollowood:2002zk}, where 
the terminology  \emph{elliptic spin Calogero-Moser
systems} was adopted.
In addition, the 4d S-duality group has a rather rigid structure that 
can be read from the geometry. It corresponds to the mapping class group of the punctured torus. Its  faithful action is generated by modular transformations acting on the total gauge coupling, together with the permutations associated with the affine Weyl symmetry of the necklace and the exchanges of the punctures. We refer the reader to \cite{Halmagyi:2004ju,Amariti:2016hlj} for more details on the S-duality group for the necklaces at hand here.

Observe that the Bethe Ansatz approach was already applied to such models (and more generically to 
4d $\mathcal{N}=1$ toric gauge theories) in \cite{GonzalezLezcano:2019nca,Lanir:2019abx, Benini:2020gjh} where the large $N$ contribution was evaluated. 
The vacua found in these references are  built by assigning the same HL solution of the $\mathcal{N}=4$ $\SU(N)$ case to each gauge node. 
However, new vacua are necessary in order to conjecture a correspondence between the discrete solutions of the Bethe Ansatz equations and the vacua of the elliptic spin Calogero-Moser systems.

A constructive approach to find the solutions to the BAEs for the necklaces can be adopted, elaborating on the results of \cite{Amariti:2025vjd}. In this reference an analytic control of the BAEs was possible for the $\SU(2) \times \SU(2)$ conifold theory.
In that case the discrete solutions were found, similarly to the cases of $\mathcal{N}=4$ SYM  $\SU(2)$ and $\SU(3)$ in \cite{Benini:2021ano}, by solving the BAEs analytically through a proper factorization of the Jacobi theta functions.
Indeed, as we observe below, such discrete solutions survive also in the simplest $\mathcal{N}=2$ necklace corresponding to the 
$\mathbb{C}^3/\Z_2$ case. 
In this case only three out of the six solutions  
found in \cite{Amariti:2025vjd}
are HL solutions in the sense spelled out in \cite{Lanir:2019abx}.
The other three solutions can be obtained from the three HL ones as described in \cite{Amariti:2025vjd}.
They are  obtained by acting on the HL ones by a rank-dependent shift on the holonomies of one of the two gauge factor (in the case at hand,   $N=2$, a semi-integer shift). This choice of the gauge factor is actually immanent, because the two  are equivalent due to the $\mathbb{Z}_2$ symmetry of the quiver.
The six solutions of the BAEs  are in agreement with the six massive vacua of the elliptic spin Calogero-Moser system studied in \cite{Hollowood:2002zk}.
Furthermore,
the six solutions do not reproduce the full SCI and indeed in the $\mathbb{C}^3/\Z_2$ case 
massless vacua for the spin Calogero-Moser system are expected\footnote{On the other hand such  six solutions reproduce the index in the case of the $\SU(2) \times \SU(2)$ conifold, because the Klebanov-Witten deformations lift all the flat directions.} from the analysis of
\cite{Hollowood:2002zk}.

In this paper we find general solutions of the BAEs of 4d $\SU(N)^K $ $\mathcal{N}=2$ necklaces,
with $N$ and $K$ arbitrary. 
We obtain a classification for the solutions analogous to the HL ones found for the $\SU(N)$ $\mathcal{N}=4$ SYM case in the literature\footnote{In analogy with the case of $\SU(N)$ $\mathcal{N}=4$ SYM 
we cannot exclude the existence of other solutions, for example solutions that can sporadically emerge for fixed values of  $N$. The solutions found here on the other hand can be defined and hold for generic values of $N$ and $K$ and have a well defined behavior under the action of the S-duality group.}.
As discussed above, only a subset of the solutions found here are actually HL  solutions, and they were already found  
in \cite{Lanir:2019abx}. We denoted such solutions as \emph{progenitors} below.
The new solutions, referred to as \emph{descendant} solutions below, are obtained, analogously to the $K=2$ case spelled out above,
by acting on the HL ones with  rank-dependent shifts.
The  picture that emerges from our classification  
suggests a map
between the number of inequivalent solutions to the BAEs
and the number of massive vacua of the spin Calogero-Moser system.
This extends the results found in the case of $\SU(N)$ $\mathcal{N}=4$ SYM, that corresponds to the case $K=1$ in the necklaces at hand. 
Beyond the numerology, we find that the transformations relating the progenitors and the descendant solutions realize the expected generators of the S-duality group.
Concretely, the progenitor solutions discussed in \cite{Lanir:2019abx} sit in an orbit under the action of modular transformations, while shifts and permutations realize the new families of descendant solutions that we have obtained here.

The paper is organized as follows.
In Section \ref{necklacereview} we review  the necklaces considered in the body of the papers. We first  provide their type IIA brane picture in sub-section \ref{theorydata}. Then we give an extensive discussion of  the S-duality group in sub-section \ref{Sduality}. We conclude our review in sub-section \ref{N1*def}  by studying the  $\mathcal{N}=1^*$ massive deformations and associated elliptic spin Calogero-Moser system, focusing on the counting of vacua and on their behavior under S-duality.
In Section \ref{secBethe} we study  the BAEs for these necklaces. In order to fix the terminology in sub-section \ref{SCIsec} we review some basic aspects of the superconformal index and in 
sub-section  \ref{BAreview} of the BA approach.
In sub-section \ref{secsol} we provide the solutions of the BAEs for the necklaces considered in this paper. First we review the known HL progenitor solutions already appeared in the literature.
Then we construct the new descendant solutions along the arguments introduced above and we explicitly prove that they solve the BAEs. 
In Section \ref{newsolprop} we elaborate on various aspects of the solutions of the BAEs and on the evaluation of the index.
We start in sub-section \ref{GHLcounting} by counting the number of inequivalent progenitor and descendant solutions to the BAEs. 
For concreteness we also give the explicit realization of a simple example, corresponding to necklace with three $\SU(3)$ gauge groups.
In sub-section \ref{M1K} we study the action of the S-duality group on the inequivalent solutions and again we give the explicit realization for the case 
with three $\SU(3)$ gauge groups.
In sub-section \ref{subsec:countingspin} we extend the first part of the Conjecture \ref{Conjecture} to the necklaces considered here. Indeed we provide evidences in favor of 
a correspondence between the inequivalent progenitor and descendant solutions and the massive vacua of the 
elliptic spin Calogero-Moser integrable system. 
We also evaluate the contributions of the solutions to the SCI of the necklace theory.
In sub-section \ref{finiteN} we comment on the finite $N$ regime for 
the necklace with two gauge nodes, showing that for $\SU(2)\times \SU(2)$ we cannot reproduce the index even if we include the contribution of the new descendant solutions in addition to the one of the original HL progenitors. 
In sub-section \ref{largeN} we compute the large $N$ contribution (for general necklaces with $K$ nodes) of the descendant solutions to the SCI.
In Section \ref{secconc} we summarize our findings and provide some speculative discussion on future lines of research.
In Appendix \ref{AppN=1*} further details on the $N=1^*$ $\SU(N)$ SYM case are collected.

\section{\texorpdfstring{$\mathcal{N}=2$ necklace theories}{}}
\label{necklacereview}

We start by reviewing the fundamental aspects of the class of 4d $\mathcal{N} = 2$ Lagrangian quiver gauge theories under consideration. 
We start by describing the model, then we discuss the action of the S-duality group and finally we study its $\mathcal{N}=1^*$ mass deformations.

\subsection{Description of the model}
\label{theorydata}

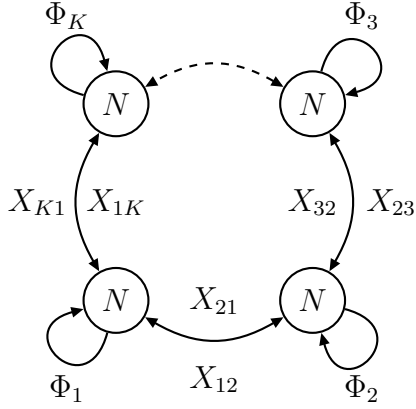
\begin{figure}
	\centering
	\begin{tikzpicture}[
		scale=1.3,
        box/.style={rectangle, draw, thick},
        ]
        \pgfmathsetmacro{\x}{1}
        \pgfmathsetmacro{\y}{2}
		\pgfmathsetmacro{\h}{0.8}
        % Nodes
        \node[,circle, draw, thick] (1) at (-\x, 0) {$N$};
        \node[circle, draw, thick] (2) at (\x, 0) {$N$};
        \node[circle, draw, thick] (3) at (\x, \y) {$N$};
        \node[circle, draw, thick] (4) at (-\x, \y) {$N$};
		\node (11) at (-\x-0.5, 0.9+\y) {$\Phi_K$};
		\node (22) at (\x+0.5, 0.9+\y) {$\Phi_3$};
		\node (33) at (\x+0.5,-0.9) {$\Phi_2$};
		\node (44) at (-\x-0.5,-0.9) {$\Phi_1$};
        % Bifundamentals
        \draw[<->,  thick, >={Triangle[angle=45:5pt]}] (1) to[out=-30, in=-150, looseness=1.1] (2);
		\node at (0, 0) {$X_{21}$};
		\node at (0, -\h) {$X_{12}$};
        \draw[<->,  thick, >={Triangle[angle=45:5pt]}] (2) to[out=60, in=-60, looseness=1.1] (3);
		\node at (\x, \y/2) {$X_{32}$};
		\node at (\x+\h, \y/2) {$X_{23}$};
        \draw[<->,  thick, dashed, >={Triangle[angle=45:5pt]}] (3) to[out=150, in=30, looseness=1.1] (4);
		\node at (-\x, \y/2) {$X_{1K}$};
		\node at (-\x-\h, \y/2) {$X_{K1}$};
        \draw[<->,  thick, >={Triangle[angle=45:5pt]}] (4) to[out=-120, in=120, looseness=1.1] (1);
		% Adjoints
        \draw[->,  thick, >={Triangle[angle=45:5pt]}, looseness=6] (1) to[out=255, in=195] (1);
        \draw[->,  thick, >={Triangle[angle=45:5pt]}, looseness=6] (2) to[out=345, in=285] (2);
        \draw[->,  thick, >={Triangle[angle=45:5pt]}, looseness=6] (3) to[out=75, in=15] (3);
        \draw[->,  thick, >={Triangle[angle=45:5pt]}, looseness=6] (4) to[out=165, in=105] (4);
    \end{tikzpicture}
	\caption{Quiver diagram of the $(N,K)$ necklace theory in $\mathcal{N} = 1$ formalism.}
	\label{KnecklaceN}
\end{figure}

The $\mathcal{N}=2$ necklace theories are described by a necklace-shaped quiver with $K$ nodes and with links (in $\mathcal{N} = 2$ language) connecting consecutive nodes among themselves, as depicted in Figure \ref{KnecklaceN}. Each theory of this class is specified by a pair of positive integers $(N,K)$ which completely determines the symmetries and the matter content of the theory. A $\mathrm{SU}(N)$ gauge group sits at each node and hypermultiplets $\Xi_a$ are associated with the links. The full gauge symmetry group is 
\begin{equation}
	G = \frac{\SU(N)^K \times \mathrm{U}(1)}{\mathbb{Z}_N}\,,
\end{equation}
with the $\mathbb{Z}_N$ diagonally embedded in all the $\SU(N)_a$ factors. The $\mathrm{U}(1)$ factor can be neglected as it decouples from the dynamics in the IR\footnote{Notice however that this $\mathrm{U}(1)$ plays a central role in the choice of global properties of the gauge group \cite{Amariti:2016hlj}.}.

In $\mathcal{N} = 1$ language, the matter content is characterized by pairs of bifundamental fields $\Xi_a = (X_{a,a+1}$, $X_{a+1, a})$ connecting consecutive nodes, and adjoint matter fields $\Phi_{a}$ arising from the $\mathcal{N} = 2$ vector multiplet, located at each node and interacting via a superpotential
\begin{equation}
	\label{Wneck}
	W = \sqrt{2}\,\sum_{a=1}^K \mathrm{Tr} 
	\left( 
		X_{a,a+1}\Phi_{a+1}X_{a+1,a} - 
		X_{a+1,a}\Phi_{a}X_{a,a+1}
		\right).
\end{equation}

Such theories have a long history in the literature. They can be obtained in type IIB string theory by placing a stack of $N$ D3 branes at an elliptic $A_{K-1}$ singularity \cite{Douglas:1996sw}. The $(N,1)$ case corresponds to $\mathcal{N} = 4$ SYM. All the others can be constructed via $\mathcal{N} = 2$ orbifold $\mathbb{C}^2/\mathbb{Z}_K \times \mathbb{C}$ and admit an holographic interpretation in terms of string theory on $\mathrm{AdS}_5 \times S^5/\mathbb{Z}_K$ background \cite{Kachru:1998ys}, sharing many properties with their parent.

Via T-duality it is also possible to engineer such family of models in type IIA string theory. The setup consists of $N$ D4 branes suspended between $K$ NS5 branes on a circle, as first discussed in \cite{Witten:1997sc}. This setup is more convenient for our purposes, as it makes the non-perturbative properties of the theory appear more natural. Indeed, by lifting the brane picture to M-theory, we can naturally identify the duality group of the theory acting on the complexified gauge couplings
\begin{equation}
	\tau_a = \frac{4 \pi \mi}{g_a^2} + \frac{\vartheta_a}{2\pi}\,, \qquad a = 1,...,K\,,
\end{equation}
as the mapping class group of a torus with $K$ punctures, $\mathcal{M} (1,K)$ \cite{Witten:1997sc, Halmagyi:2004ju}. We briefly review the basic details.

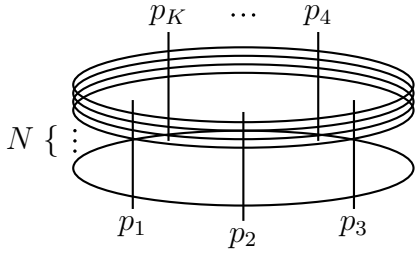
\begin{figure}
\centering
\begin{tikzpicture}[scale=0.45]

% Ellissi superiori
\draw[black, line width=0.8pt] (0,0.45) ellipse (5.0 and 1.10);
\draw[black, line width=0.8pt] (0,0.20) ellipse (5.0 and 1.10);
\draw[black, line width=0.8pt] (0,-0.05) ellipse (5.0 and 1.10);
\draw[black, line width=0.8pt] (0,-0.30) ellipse (5.0 and 1.10);

% Ellisse inferiore
\draw[black, line width=0.8pt] (0,-2) ellipse (5.0 and 1.10);

% Segmenti verticali
\draw[black, line width=0.8pt] (-3.25,-3.20) -- (-3.25,-0.0);
\draw[black, line width=0.8pt] (0.00,-3.55) -- (0.00,-0.35);
\draw[black, line width=0.8pt] (3.25,-3.20) -- (3.25,-0.0);

\draw[black, line width=0.8pt] (-2.20,-1.2) -- (-2.20,2);
\draw[black, line width=0.8pt] (2.20,-1.2) -- (2.20,2);

% Puntini superiori
\fill[black] (-0.3,2.5) circle (0.05);
\fill[black] (0.0,2.5) circle (0.05);
\fill[black] (0.3,2.5) circle (0.05);

% Puntini laterali presso N
\fill[black] (-4.95,-0.9) circle (0.05);
\fill[black] (-4.95,-1.2) circle (0.05);
\fill[black] (-4.95,-1.5) circle (0.05);

% Etichette
\node at (-6.2,-1.2) {$N\;\{$};

\node at (-3.25,-3.70) {$p_1$};
\node at (0.00,-4.05)  {$p_2$};
\node at (3.25,-3.70)  {$p_3$};

\node at (-2.20,2.5)  {$p_K$};
\node at (2.20,2.5)   {$p_4$};

\end{tikzpicture}
\caption{Brane configuration in IIA string theory. We have a stack of $N$ D4-branes wrapping around $x^6$ and $K$ NS5-branes at fixed positions $p_1,\dots,p_K$ along $x^6$.}
\label{branesetup}
\end{figure}

We consider type IIA in $\mathbb{R}^{1,9}$, labeled by coordinates $x^0,\dots,x^9$, and we compactify the coordinate $x^6$ on a circle.
We study the setup depicted in Figure \ref{branesetup}. We introduce $K$ NS5 branes extended along $x^0,x^1,\dots, x^5$ and located at fixed positions $p_i$ along $x^6$.  Suspended between consecutive NS5 branes, we introduce stacks of $N$ D4 branes, whose world-volumes extends along directions $x^0,\, x^1, \dots, x^3$ and wraps $x^6$, while lies at a fixed position in the coordinates $x^4,\, x^5$ along which the NS5 are extended. The setup preserves $\mathcal{N}=2$ supersymmetry and the worldvolume theory emerging along $(x^0,\cdots,x^3)$ engineers the $\mathcal{N}=2$ 4d field theory described by the necklace quiver models previously introduced.

We introduce the holomorphic coordinate $v = x^4 + \mi x^5 \in \mathbb{C}$ to conveniently parameterize the position of the D4s. Relative positions of consecutive stacks of D4 parameterize bare mass deformations for the hypermultiplets of the theory, up to an appropriate constant prefactor
\begin{equation}
	m_a \sim v_{a+1} - v_a\,.
\end{equation}

The periodicity along $x^6 \sim x^6 + 2 \pi L$ constrains the $K$ mass parameters as $\sum_{a=1}^K m_a = 0$. As discussed in \cite{Witten:1997sc}, one can have free mass parameters $\sum_{a=1}^K m_a = m$ by considering a non-trivial bundle of the geometry spanned by $(x^4,x^5,x^6)\in \mathbb{C} \times S^1$. This is precisely achieved starting from the covering space $\mathbb{C}\times \mathbb{R}$ and considering the non-trivial $\mathbb{C}$-bundle over $S^1$ obtained by the quotient
\begin{equation}
	\begin{cases}
		x^6 \mapsto x^6 + 2\pi L\\
		v \mapsto v + m
	\end{cases}.
\end{equation} 

The distance between consecutive NS5s defines the gauge couplings $g_a^2$ of each gauge node
\begin{equation}
	\frac{1}{g_a^2} = \frac{x^6_{a+1} - x^6_{a}}{8\pi g_s L}\,.
\end{equation}
The lift to M-theory is natural from this perspective.
If we introduce the $\theta$-angle $\theta_a$ and the complexified gauge couplings $\tau_a$
\begin{equation}
	\tau_a = \frac{4 \pi \mi}{g_a^2} + \frac{\theta_a}{2 \pi}\,,
\end{equation}
we can introduce a new periodic variable $x^{10} \sim x^{10} + 2\pi R$ and define $p = x^6 + \mi x^{10}$ such that
\begin{equation}
	\tau_a = \frac{\mi (p_{a+1} - p_{a})}{2\pi R}\,.
\end{equation}
Therefore, we lift the setup to M-theory by considering the theory on $\mathbb{R}^{1,9}\times S^1$.\footnote{Notice that $g_a^2$ is insensitive to a simultaneous rescaling of $g_s$ and the relative $x_6$ positions of the NS5s. Thus we are allowed to take the M-theory limit $g_s \to \infty$ without affecting the 4d description, by simultaneously rescaling also the relative $x^6$ distance between consecutive NS5s, as commented in \cite{Witten:1997sc}.} In order to consider a non-zero $\tau = \frac{4 \pi \mi}{g^2} + \frac{\theta}{2\pi} =\sum_{a=1}^K \tau_a$ the geometry of $(v,p)$ is only locally a product $\mathbb{C} \times \mathbb{T}$, with $\mathbb{T}$ a torus, while globally we have the non-trivial $\mathbb{C}$-bundle $E$ over the genus one Riemann surface $\mathbb{T}_\tau$, with complex structure $\tau$, defined by the quotient
\begin{equation}
	\begin{cases}
		v \mapsto v + m \\
		x^6 \mapsto x^6 + 2\pi L \\
		x^{10} \mapsto x^{10} + \theta R
	\end{cases},
\end{equation} 
together with the subsequent quotient $x^{10} \sim x^{10} + 2\pi R$.

Moving to M-theory, the configuration of intersecting D4s and NS5s becomes smooth and it lifts to a unique M5-brane, incorporating both D4 and NS5 worldvolumes into a more complicated one $\mathbb{R}^{1,3} \times \Sigma_g$, with $\Sigma_g \subset E$ being a genus $g = K(N - 1) + 1$ Riemann surface described by the Seiberg-Witten curve presented in \cite{Witten:1997sc}. The rank of the gauge group emerging in the low-energy theory upon compactification along $\Sigma_g$ is $g$, precisely matching the four dimensional necklace model, indeed the dimension of the Coulomb branch of $\mathrm{SU}(N)^{K} \times \mathrm{U}(1)$ is $K(N-1) + 1$. 

The curve $\Sigma_g$ is a branched $N$ fold cover of $\mathbb{T}_\tau$ spanned by $s$, where the N fold covers arise due to the M-theory lift of the $N$ D4 branes, wrapping $(x^6,x^{10})$. At the same time, the $K$ positions $p_a$ of the NS5 define $K$ punctures on $\mathbb{T}_\tau$.

\subsection{The S-duality group}
\label{Sduality}

It is now immediate to derive the S-duality group of the $\mathcal{N}=2$ low-energy 4d theories. Let us consider the moduli space  $\mathcal{M}_{1,K}$ of a torus with $K$ punctures $\mathbb{T}_{1,K}$. A point $w(\tau_a)$ in the moduli space specifies all the details about the couplings $g_a, \theta_a$ of the theory. Upon performing an isomorphism of $\mathbb{T}_{1,K}$ the actual coupling of the theory $w(\tau_a')$ is left invariant. However, the typical definition of the low-energy 4d theory is in terms of $\tau_a$. Therefore, the two theories will be related by a duality group transformations $\tau_a \mapsto \tau_a'$, acting as a symmetry from the point of view of local operators of the theory. This is translated, roughly speaking, into a multi-value behavior of the inverse function $\tau(w)$. The duality group is then classified by the fundamental group $\pi_1(\mathcal{M}_{1,K})$, namely the mapping class group of a genus one Riemann surface with $K$ punctures, that we will denote $\mathcal{M}(1,K)$.

In the simplest case of $K=1$, $\mathcal{M}(1,1) = \mathrm{SL}(2,\mathbb{Z})$, which is the S-duality group of $\mathcal{N} = 4 $ SYM. 
For generic $K$, the group is more complicated, as it includes modular transformations acting on each gauge coupling $\tau_a$, together with permutations and period shifts of indistinguishable punctures on the torus. Indeed, moving NS5 branes along the torus cycles or exchanging them leaves the configuration invariant.

An explicit definition of the mapping class group $\mathcal{M}({1,K})$ can be found in \cite{Halmagyi:2004ju}. We denote $\sum_{a=1}^K \tau_a  = \tau$ and introduce the generators:
\begin{equation}
\begin{aligned}
	\label{eq:MCG}
		S\,:\;&(\tau;p_1,...,p_K)\mapsto \left(-\tfrac{1}{\tau};\tfrac{p_1}{\tau},...,\tfrac{p_K}{\tau}\right)\\
		T\,:\;&(\tau;p_1,...,p_K)\mapsto (\tau+1;p_1,...,p_K)\\
		T_a\,:\;&(\tau;p_1,...,p_a,...,p_K)\mapsto (\tau;p_1,...,p_{a}+1,...,p_K)\\
		t_a\,:\;&(\tau;p_1,...,p_a,...,p_K)\mapsto (\tau;p_1,...,p_a+\tau,..., p_K)\\
		\sigma_a\,:\;&(\tau;p_1,...,p_a,p_{a+1},...,p_K)\mapsto (\tau;p_1,...,p_{a+1},p_a,...,p_K)\\
		\sigma_K\,:\;&(\tau;p_1,...,p_K)\mapsto (\tau;p_K - \tau,p_2,...,p_{K-1},p_1+\tau)\,.
\end{aligned}
\end{equation}
We further impose the following condition 
\begin{equation}
	\label{eq:MCGcondition}
	(\tau;p_1,...,p_K) \sim (\tau;p_1 + c,..., p_K + c)\,, \quad \quad c \in \mathbb{C}\,,
\end{equation}
which reflects the fact that only relative differences among the punctures are physically meaningful. The set of generators introduced above is redundant. There are indeed many relations among such generators. By introducing $\omega$
\begin{equation}
	\omega = t_1 \sigma_{K} \sigma_{K-1}...\sigma_{2}\,,
\end{equation}
which acts on the punctures as 
\begin{equation}
	(\tau;p_1,...,p_K)\mapsto (\tau;p_2,...,p_K,p_1+\tau)
\end{equation}
and satisfies the cyclic $\mathbb{Z}_K$ condition $\omega^K = 1$ due to \eqref{eq:MCGcondition}, one can easily see that all the generators listed above can be obtained by the iterated action of elements of the set $\{S,T,\sigma_1,\omega\}$. Thus $\mathcal{M}(1,K)$ generalizes the $\mathrm{SL}(2,\mathbb{Z})$ group to a larger one that also contains $\mathbb{Z}_K$ and the affine Weyl group $\hat{A}_{K-1}$.

\subsection{\texorpdfstring{$\mathcal{N}=1^*$ massive deformations}{}}
\label{N1*def}

Necklace models admit a family of $\mathcal{N} = 1^*$ massive deformations, studied in \cite{Dorey:2001qj},  which can be derived via deformation of the original $(N,K)$ theory by giving masses to the hypermultiplets as well as the adjoint fields of the theory, via the $\mathcal{N} = 1$ superpotential
\begin{equation}
    \delta W = \sum_{a=1}^K \left(\mu_a X_{a,a+1}X_{a+1,a} + m_a \Phi_{a,a}^2\right). 
\end{equation}
Such deformations allow for a rich structure of vacua and phases of the theory.
The quantum structure of the vacua can be studied by first considering the soft $\mathcal{N} = 2^*$ deformation of the theory, obtained by giving masses only to the hypers $\mu_a \neq 0$ as discussed in \cite{Dorey:2001qj}, and analyzing the special singular points in the Coulomb branch of the theory where the aforementioned Seiberg-Witten curve $\Sigma_g$ undergoes maximal degeneracy to a genus one surface. 
Equivalently, the vacua of the theory can be studied via the integrable system associated to the Seiberg-Witten curve, which in this case defines a spin generalization of elliptic Calogero-Moser systems \cite{Nekrasov:1995nq, Dorey:2001qj}.
The classification of vacua and phases generalizes the one of $\mathcal{N} = 1^*$ SYM. From the semiclassical analysis, F-flatness modulo $\mathrm{SL}(N,\mathbb{C})^K$ transformations classifies vacua in terms of partitions of ordered set of $K N$ elements:
\begin{equation}
    \begin{aligned}
        &\{1,2,...,K,1,2,...,K,......,1,...,K\} \;\longmapsto \; \{\mathcal{A}_1,...\mathcal{A}_n\}\,,
        \\
        \mathcal{A}_j &= \{i_{j-1}+1,...,i_j\}
        \,, \qquad
        \dim\mathcal{A}_j = \ell_j
        \,, \qquad 
        \sum_{j=1}^n \ell_j = K N
    \end{aligned}
\end{equation}
where each $\ell_j$ is associated with the embedding of a $\ell_r$ dimensional representation of $\mathrm{SU}(2)$ block within $\mathrm{SU}(N)^K$. The allowed partitions must satisfy the condition
\begin{equation}
\{i_1,...i_n\} \supset \{1,2,...,K\}.
\label{eq:partconst}
\end{equation}
Additionally, permutations of the subsets $\mathcal{A}_j$ produce an equivalent, indistinguishable, partition. This fact reflects the action of the Weyl group of $\mathrm{SU}(N)^K$ on the vacua. As a consequence of condition \eqref{eq:partconst}, the number of subsets $\mathcal{A}_i$ satisfies $n\geq K$. We summarize the classification of vacua \cite{Dorey:2001qj}:
\begin{itemize}
    \item Generic partitions may leave unbroken abelian factors in the gauge group. In such a case they are associated with branches of complex dimensional massless Coulomb vacua, in analogy with the discussion above for $\mathcal{N}=1^*$ SYM.
    \item The minimal partitions $n = K$ are of the form $\{\mathcal{A}_1,...,\mathcal{A}_K\}$. These partitions leave an empty unbroken gauge group. For this reason, they are associated with a massive Higgs vacuum. From a simple combinatorial argument, thanks to the ordered structure of the set, one can easily see that there are $N^{K-1}$ inequivalent partitions and thus $N^{K-1}$ Higgs vacua.
     In contrast, the maximal partition $n = NK$ contains a single element in each subset $\mathcal{A}_i$. Thus, we have $N$ copies of identical subsets $\mathcal{A}_i = \{i\}$. In this case, the whole $\mathrm{SU}(N)^K$ gauge group is left unbroken and such partition is then associated with massive confining vacua. In this case the low-energy theory is pure $\mathcal{N} = 1$ SYM, thus we have $N$ confining vacua for each gauge group, $N^K$.
     More generally, for each divisor of $N = p q$ we have an intermediate situation, where the partition is of the form 
    \begin{equation*}
        \left\{
        \mathcal{A}_1,\dots,\mathcal{A}_1,
        \mathcal{A}_2,\dots,\mathcal{A}_2,
        \dots\dots,
        \mathcal{A}_K,\dots,\mathcal{A}_K
        \right\},
    \end{equation*}
    where each subset repeats $p$ times. The associated partition is an Higgs vacuum for a $\mathrm{SU}(q)^K$ gauge group and leaves an unbroken $\mathrm{SU}(p)^K$ gauge groups, which confines. The number of massive vacua is then $p^K q^{K-1} = N^{K-1} p$. 
\end{itemize}

All in all, the total number of massive vacua of $\mathcal{N} = 1^*$ necklace models generalizes the counting of $\mathcal{N} =1^*$ SYM, and it is given by
\begin{equation}
    \sum_{p|N} N^{K-1} p = N^{K-1} \sigma_1(N).
\end{equation}
We will see that the very same counting of massive vacua will also be reproduced by our set of solutions to the BAEs of the $\mathcal{N} = 2$ models.

The massive vacua group into families associated with distinct phases of the theory. The presence of bifundamental matter breaks the center 1-form symmetry $\mathbb{Z}_N^{K}$ to a diagonal $(\mathbb{Z}_N)_D$. For fixed gauge group the phases of the theory are in one-to-one correspondence with the order-$N$ sublattices of $(\mathbb{Z}_N)_D \times (\mathbb{Z}_N)_D$ of mutually local electric and magnetic line probes of total electric charge $e$ and magnetic charge $m$, associated to distinct global variants of the theory \cite{Aharony:2013hda}. In complete analogy with $\mathcal{N} =1^*$ SYM there are $\sigma_1(N)$ distinct phases, now with $N^{K-1}$ vacua in each family. 

This structure reflects the action of the S-duality group of the original $\mathcal{N} = 2$ models on the lattices of line operators \cite{Amariti:2016hlj}. Mutually local line operators are specified by a vector of charges $(e_1,...,e_K;m_1,...,m_K)$. As a consequence of the presence of bifundamental matter, the DSZ condition for mutually local lines forces $m_i = m_{i+1}$ for each gauge node of the theory. The line probes allowed in different global variants are thus organized in terms of sublattices of $(\mathbb{Z}_N)_D \times (\mathbb{Z}_N)_D$ for the 1-form symmetry of residual $(\mathbb{Z}_N)_D$, specified by the pair $(e,m)$ with total electric charge $e = \sum_{a=1}^K e_a $. Line operators transform according to the orbits of the S-duality group $\mathcal{M}(1,K)$. The $\mathrm{SL}(2,\mathbb{Z})$ subgroup acts only on lattices $(e,m)$ of the total electric charge $e$, and magnetic charge $m$, while it is insensitive to the details of the individual charges $(e_1,...,e_K,m)$ of a line. The other generators of $\mathcal{M}(1,K)$ act as automorphisms of such lattices, via transformations on the lines, preserving the total charge $(e,m)$. 
The vacua of the deformed theory ``spontaneously break'' the S-duality group, meaning that they inherit the very same action of $\mathcal{M}(1,K)$ on the vacua. Focusing on the massive ones, the $\mathrm{SL}(2,\mathbb{Z})$ subgroup acts by transforming the $\sigma_1(N)$ distinct families among themselves, while the remaining generators act on the $N^{K-1}$ vacua within each family.

\section{\texorpdfstring{Bethe Ansatz approach for $\mathcal{N}=2$ necklace theories}{}}
\label{secBethe}

In this section we apply of the Bethe Ansatz approach to the $\mathcal{N}=2$ necklace theories.
First, we briefly review the main aspects of this technique, showing the already known results about the limit case of $K=1$, \emph{i.e.} $\mathcal{N}=4$ SYM.
Then, we provide a generalization to the case of any $K$ of such results.

\subsection{The Superconformal Index}
\label{SCIsec}

The superconformal index (SCI) can be defined for any supersymmetric theory enjoying an $\UU(1)$ $R$-symmetry and can have different realizations depending on the amount of supersymmetry.
Even though our focus are theories with $\mathcal{N}=2$ supersymmetry, we will discuss the index in the $\mathcal{N}=1$ language where the Bethe Ansatz approach is naturally formulated.

The SCI counts protected local operators in short representations of the superconformal (sub-)algebra $\mathfrak{su}(2,2|1)$.
Chosen a complex supercharge $\mathcal{Q}$, the SCI is defined as a Witten index of the theory on $S^3$, refined by the commutant of the supercharge $\mathcal{Q}$:
\begin{equation}
    \label{SCIdef}
    \mathcal{I} = \Tr_{\mathcal{H}_{S^3}} (-1)^F 
    \ee^{-\beta \{\mathcal{Q},\mathcal{Q}^\dag\} }
    p^{J_1+\frac{r}{2}}q^{J_2+\frac{r}{2}}v_{\mu}^{Q_{\mu}}\,.
\end{equation}
Here, $J_{1,2}$ are the angular momenta of $S^3$, $r$ is the $\mathrm{U}(1)_r$ $R$ charge, $Q_\mu$ are the generators of any global symmetries of the theory, and $p$, $q$ and $v_\mu$ are fugacities associated to each symmetry.
The SCI receives contributions only from states that are annihilated by $\mathcal{Q}$.

In the case of a gauge theory the SCI admits a representation in terms of an elliptic hypergeometric integral \cite{Dolan:2008qi}.
This result can be obtained either by exploiting invariance under small deformations and through an evaluation at weak coupling \cite{Kinney:2005ej, Sundborg:1999ue,Aharony:2003sx}, or using localization techniques on the Euclidean supersymmetric partition function, on $S^1 \times S^3$, which is related to the SCI (up to an overall phase due to the supersymmetric Casimir energy) \cite{Pestun:2007rz,Closset:2013vra,Assel:2014paa}.
With both approaches the SCI can be recast as a matrix integral 
\begin{equation}
    \label{SCInt}
    \mathcal{I}(v;p,q) = 
    \kappa \oint_{\numberset{T}^{\mathrm{rk}\,G}}
    \prod_{i=1}^{\mathrm{rk}\,G} \frac{\dd{z_i}}{2 \pi \mi z_i}\,
    \mathcal{Z}(z,v;p,q)\,,
\end{equation}
where
\begin{equation}
    \label{integrand}
    \kappa = 
    \frac{ (p;p)_\infty^{\mathrm{rk}\,G}(q;q)_{\infty}^{\mathrm{rk}\,G} }{|\mathcal{W}_G|}\,,
    \qquad
    \mathcal{Z}(z,v;p,q) = 
    \frac
    {\prod_{b=1}^{n_\chi} \prod_{\rho_b \in \mathfrak{R}_b} \Gamma_e\left( (pq)^{\frac{r_b}{2}} z^{\rho_b} v^{\omega_b}; p, q \right)}
    {\prod_{b=1}^{n_V}\prod_{\alpha_b \in \Delta_b} \Gamma_e\left(z^{\alpha_b};p, q\right)}\,.
\end{equation}
The integral runs over the gauge variables $z_i=e^{2\pi \mi u_i}$, with $i=1,\dots,\mathrm{rk}\,G$, where $u_i$ are the gauge holonomies of the group $G$.
The numerator of $\mathcal{Z}$ contains the contribution of the matter content consisting of $n_{\chi}$ chirals $\Phi_b$ with R-charges $r_b$, transforming in representations $\mathfrak{R}_b$ of the gauge group $G$, with weights $\rho_b$, and in representation $\mathfrak{R}_F$ of the flavor group $G_F$, with weights $\omega_b$\footnote{We define $z^{\rho} = \prod_{i = 1}^{\mathrm{rk}\,G} z_i^{\rho_i}$, and analogously $v^\omega$.}.
The denominator consists of the contribution of the $n_b$ vector multiplets of the theory, with $\alpha_b$ being the roots of the group $G$. 
The integral contour is the maximal torus $\numberset{T}^{\mathrm{rk}\,G}$ of $G$ and $|\mathcal{W}_G|$ is the cardinality of the Weyl group of $G$.
The $q$-Pochhammer symbol $(q;q)_\infty$ and the elliptic gamma function $\Gamma_e(z;p,q)$ are 
defined as
\begin{equation}
	\label{pochhegamma}
	(p;q)_{\infty} = 
	\prod_{n=0}^{\infty} \left( 1-p q^n \right)\,,
	\qquad
	\Gamma_e(z;p,q) =
	\prod_{m,n=0}^{\infty} \frac{ 1 - p^{m+1}q^{n+1}z^{-1} }{ 1 - p^{m}q^{n}z }\,.
\end{equation}
In what follows, it will be useful to introduce the chemical potentials
\begin{equation}
	\label{fugchem}
	v_\mu=\ee^{2 \pi \mi \xi_\mu}, \quad
	p = \ee^{2 \pi \mi \sigma}, \quad
	q = \ee^{2 \pi \mi \tau}, \quad
	y_a = v_a (pq)^{\frac{r_a}{2}} = \ee^{2\pi \mi \Delta_a},
\end{equation}
with 
\begin{equation}
	\label{DeltaDef}
	\Delta_a = \omega_a(\xi) + r_a \frac{\tau + \sigma}{2}.
\end{equation}
The superconformal index for the $\mathcal{N} = 2$ necklace with $K$ $\SU(N)$ gauge nodes is given by 
\begin{equation}
	\label{SCIneck}
	\begin{aligned}
		\mathcal{I} &= 
		\left(\frac{ (p;p)_\infty^{N-1}(q;q)_{\infty}^{N-1} }{N!}\right)^{\!K}
		\prod_{a=1}^K \Gamma_e\left( y_{a,a} \right)^{N-1}
		\oint_{\mathbb{T}^{K(N-1)}}
		\left(\,
		\prod_{a=1}^{K} \prod_{i=1}^{N-1}
		\frac{\dd{z_i^{(a)}}}{2 \pi \mi z_i^{(a)}} 
		\,\right) \\
		&
		\prod_{a=1}^{K}
		\left(
		\prod_{i \ne j}^{N}
		\frac{
			\Gamma_e\left(y_{a,a} \,z_i^{(a)}/z_j^{(a)}\right)
		}{\Gamma_e\left(z_i^{(a)}/z_j^{(a)}\right)}
		\right)
		\left(
		\prod_{i, j=1}^{N}
		\Gamma_e\left(y_{a,a + 1}\, z_i^{(a)}/z_j^{(a + 1)}\right)
		\Gamma_e\left(y_{a + 1, a}\, z_i^{(a + 1)}/z_j^{(a)}\right)
		\right),
	\end{aligned}
\end{equation}
where for shortness the dependence on the fugacities $p$ and $q$ in the elliptic gamma functions is suppressed. The gauge variables on each node $a$ are constrained by the $\SU(N)$ constraint $\prod_{i=1}^N z_i^{(a)} = 1 $.
For each  node $a$, we have assigned a chemical potential $\Delta_{a,b}$ to each field transforming non-trivially under this gauge node, namely to one adjoint $\Phi_{a}$, two fundamentals $X_{a,a \pm 1}$ and two antifundamentals $X_{a \pm 1,a}$.
Each $\Delta_{a,b}$ can be expressed as a linear combination  of the chemical potentials associated with the linearly independent $\UU(1)$ factors spanning the Cartan subalgebra of the global symmetry group.
For simplicity, we keep this dependence implicit.
The invariance of the superpotential \eqref{Wneck} under flavor and $R$-symmetry then implies the following constraints on the chemical potentials
\begin{equation}
	\label{Dconstr}
	\Delta_{a,a} + 
	\Delta_{a,a\pm1} +
	\Delta_{a\pm1,a} = \tau + \sigma\,, 
	\qquad \forall\,a=1,\dots,K\,.
\end{equation}

\subsection{The Bethe Ansatz approach}
\label{BAreview}

The SCI admits an alternative representation in terms of an expansion over Bethe vacua expressed by the following general formula\footnote{The Bethe Ansatz approach holds for general $\tau\neq \sigma$. For simplicity of presentation, here and in the rest of the paper we restrict to the collinear case $\tau = \sigma$. However, we remark that many of our results are general and can be immediately extended to the non-collinear case.}
\cite{Closset:2017bse, Benini:2018mlo}:
\begin{equation}
    	\label{baf}
	\mathcal{I} = \kappa \sum_{\hat{u} \in \mathfrak{M}_\text{{BAE}}}
	\mathcal{Z}(\hat{u},\Delta;\tau)\,
	\mathcal{H}(\hat{u},\Delta;\tau)^{-1},
\end{equation}
with $\mathcal{H}(\hat{u},\Delta;\tau)$ defined in \eqref{jacobian} below, and where the sum runs over the Bethe vacua $\hat{u}\in \mathfrak{M}_\text{{BAE}}$, \textit{i.e.} solutions to a set of coupled transcendental equations referred to as Bethe Ansatz Equations (BAEs) on which the SCI localizes. Notice that we collectively denote the dependence on the chemical potentials \eqref{DeltaDef} with $\Delta$.
Equating \eqref{SCInt} and \eqref{baf} yields the 
Bethe Ansatz approach to the evaluation of the SCI, in which the computation of the index is reformulated as the problem of finding solutions to an associated Bethe problem, revealing an underlying integrable system structure emerges.
The Bethe Ansatz Equations are defined by:
\begin{equation}
	\label{baes}
	Q_i(u,\Delta;\tau)
	=1\,, \qquad
	\forall \, i=1,\dots,\text{rk}\,G\,,
\end{equation}
where we introduced the Bethe Ansatz Operators of the theory (BAOs)
\begin{equation}
	\label{basBAO}
    Q_i(u,\Delta;\tau)=
	\prod_{b=1}^{n_\chi}
	\prod_{\rho_b \in \mathfrak{R}_b}
	P\left(
		\rho_b(u) + \Delta_b; \tau
	\right)^{\rho_b^i}, \qquad
	P(u;\tau) =
    \frac{\ee^{ \pi \mi \left(u-\frac{u^2}{\tau}\right)}}
    {\theta_0(u;\tau)}\,,
\end{equation}
with $\theta_0(z;q) = (z;q)_{\infty} (q/z;q)_{\infty}$\footnote{
	This definition is written in terms of fugacities $z=\ee^{2 \pi \mi u}$ and $q=\ee^{2 \pi \mi \tau}$. However, throughout this work we will regard $\theta_0$ as a function of the chemical potentials and therefore write it as $\theta_0(u;\tau)$.
	}.
We also define the Jacobian $\mathcal{H}$
\begin{equation}
	\label{jacobian}
	\mathcal{H}(u,\Delta;\tau) = 
    \det_{ij} \left(
        \frac{1}{2 \pi \mi} 
        \frac{\partial Q_i}{\partial u_j}
    \right).
\end{equation}
In any well-defined theory with a set of non-anomalous global symmetries, the Bethe operators are elliptic, modular-invariant functions in $u$.
As a consequence, the BAEs are naturally defined on a torus $\mathbb{T}_\tau$. In particular, the holonomies $u_i$, originally defined on $S^1$, are promoted to coordinates on the torus $\mathbb{T}_\tau  =S^1 \times S^1_\tau$), and the solutions are defined on the complex torus with modular parameter $\tau$
\begin{equation}
	u \sim u+1 \sim u+\tau\,.
\end{equation}

Moreover, as proved in \cite{Benini:2018mlo}, only the solutions that are not fixed by any non-trivial element of the Weyl group of $G$ $\mathcal{W}_G$ contribute to the SCI.
The set of contributing solutions can then be written as
\begin{equation}
	\mathfrak{M}_\text{BAE}=
	\left\{\,
        \hat u \in \T_\tau^{\text{rk}\,G} :
        Q_i(\hat{u},\Delta;\tau) = 1,\,
        w \cdot \hat{u} \ne \hat{u},\,
        \forall \, i=1,\dots,\mathrm{rk}\,G,\,
        \forall \, w  \in \mathcal{W}_G \backslash \{1\}\,
    \right\}.
\end{equation}

\subsection{\texorpdfstring{Solutions to the BAEs for $\mathcal{N}=2$ theories}{}}
\label{secsol}
We now want to write the BAEs for the $\mathcal{N}=2$ necklace theory $A_{K-1}$. The BAEs of a $\SU(N)$ group can be written starting from the corresponding BAEs for the $\UU(N)$ group. They are related as follows:
\begin{equation}
Q_i^{\SU(N)}=Q_i^{\UU(N)}/Q_N^{\UU(N)} \quad \text{with } i=1,\dots, N-1. 
\end{equation}
We introduce a Lagrange multiplier $\lambda^{(a)}$ for each node $a$ to enforce the $\mathrm{\SU(N)}$, which in this language we rewrite as\footnote{Recall that each gauge holonomy is defined on a complex torus with modular parameter $\tau$.} 
\begin{equation}
	\label{suconstrTOR}
	\sum_{i=1}^N u_i^{(a)} = 0 
	\mod \Z + \tau \Z\,, \qquad
	\forall\,a=1,\dots K\,.
\end{equation}
This allows to write $Q_N^{(a)}=e^{-2\pi \mi \lambda}$. 
The BAEs \eqref{baes} for the necklace theories are
\begin{equation}
	\label{NKbaes}
	Q_{i}^{(a)} = \ee^{2 \pi \mi \lambda^{(a)}} 
	\prod_{j=1}^N
	\frac{
		P\!\left(\!\Delta_{a,a}\!+\!u_{ij}^{(a,a)};\tau\!\right)\!
		P\!\left(\!\Delta_{a,a+1}\!+\!u_{ij}^{(a,a+1)};\tau\!\right)\!
		P\!\left(\!\Delta_{a,a-1}\!+\!u_{ij}^{(a,a-1)};\tau\!\right)
	}{
		P\!\left(\!\Delta_{a,a}\!-\!u_{ij}^{(a,a)};\tau\!\right)\!
		P\!\left(\!\Delta_{a+1,a}\!-\!u_{ij}^{(a,a+1)};\tau\!\right)\!
		P\!\left(\!\Delta_{a-1,a}\!-\!u_{ij}^{(a,a-1)};\tau\!\right)
	}=1\,,
\end{equation}
where we introduced the short notation $u_{ij}^{(a,b)} = u_i^{(a)} - u_j^{(b)}$ and the chemical potentials are constraint by \eqref{Dconstr} with $\sigma=\tau$. We now introduce the Jacobi  $\theta_1(u;\tau)$ function defined as
\begin{equation}
	\label{theta1}
	\theta_1(u;\tau) = 
	\mi \left(q;q\right)_{\infty} 
	\ee^{-\pi \mi \left(u-\frac{\tau}{4}\right)}
	\theta_0(u;\tau)\,.
\end{equation}
Using the constraints on the chemical potentials \eqref{Dconstr} and the parity and quasi-double-periodicity properties of the Jacobi $\theta_1(u;\tau)$ function 
\begin{equation}
	\theta_1(u+m+n\tau;\tau) = 
	(-1)^{m+n} \ee^{- 2 \pi \mi n u} \ee^{- \pi \mi n^2 \tau}
	\theta_1(u;\tau)\,,
	\qquad
	\theta_1(-u;\tau) = - \theta_1(u;\tau)\,,
\end{equation}
we rewrite the BAEs as
\begin{equation}
	\label{NKbaesTH1}
	\begin{aligned}
		Q_i^{(a)} \!= \!
		\ee^{2 \pi \mi \lambda^{(a)}}\!
		\prod_{j=1}^N \;
		\frac{
			\theta_1\!\left(
				\Delta_{a,a} + u_{ji}^{(a,a)}
				\right)}{
			\theta_1\!\left(
				\Delta_{a,a} - u_{ji}^{(a,a)}
				\right)}
		&\frac{
			\theta_1\!\left(
				\Delta_{a+1,a}
				+u_{ji}^{(a+1,a)}
				\right)}
                {\theta_1\!\left(
				\Delta_{a,a+1} - u_{ji}^{(a+1,a)}
				\right)} 
        \frac{
			\theta_1\!\left(
				\Delta_{a-1,a}
				+u_{ji}^{(a-1,a)}
				\right)}
                {\theta_1\!\left(
				\Delta_{a,a-1} - u_{ji}^{(a-1,a)}
				\right)}
		=1,
	\end{aligned}
\end{equation}
where we redefined $\Delta_{a,a} = -\Delta_{a,a+1}-\Delta_{a+1,a}$, and we suppressed the explicit dependence on $\tau$ in $\theta_1(u;\tau)$.
The terms independent of both $i$ and the holonomies have been reabsorbed into a redefinition of the Lagrange multiplier $\lambda^{(a)}$.
Notice that for $K=1$ the equation reduce to the usual $\mathcal{N}=4$ BAEs:
\begin{equation}
	\label{SYMbaes}
	Q_i = \ee^{2 \pi \mi \lambda} 
	\prod_{\Delta \in \delta} \prod_{j=1}^N
	\frac{\theta_1(\Delta-u_i+u_j;\tau)}{\theta_1(\Delta+u_i-u_j;\tau)} = 1\,,
    \qquad
    \forall \, i=1,\dots,N\,,
\end{equation}
where $\delta = \{ \Delta_1, \Delta_2, -\Delta_1-\Delta_2 \}$.

\subsubsection*{The Hong-Liu solutions}
Due to the invariance of these equations under a common shift of the holonomies $u^{(a)}_i\mapsto u_{i}^{(a)} + \bar{u}$, a first class of solutions is given by the following \cite{Hosseini:2016cyf, Hong:2018viz,Lanir:2019abx}:
\begin{equation}
	\label{HLtor}
	\left\{m,n,r\right\}\,: \quad
	\hat{u}_{i}^{(a)}=\hat{u}_{\hat{\jmath}\hat{\kappa}}^{(a)} = 
	\bar{u} + 
	\frac{\hat{\jmath}}{m} + 
	\frac{\hat{\kappa}}{n} \left(
		\tau + \frac{r}{m}
	\right) 
	=
	\bar{u} + 
	\frac{\hat{\jmath} + \hat{\kappa} \tilde{\tau}}{m}\,, \quad \forall\,a = 1,\dots,K\,,
\end{equation}
where $\hat{\jmath}=0,\dots, m-1$ and $\hat{\kappa}=0,\dots, n-1$ parameterize the index $i=1,\dots,N$ and $m$, $n$, $r$ are integers such that $mn = N$ and $r=0,\dots,n-1$ and $\bar{u}$ is a complex number, common to all nodes enforcing the $\SU(N)$ constraint
\begin{equation}
	\label{SUconstrSYM}
	\bar{u} \; : \; \sum_{\hat{\jmath},\hat{\kappa}} \hat{u}_{\hat{\jmath} \hat{\kappa}} = 0 \mod \Z + \tau \Z\,.
\end{equation}
These solutions are referred to as Hong Liu (HL) solutions in the literature. In particular, we will refer to $\mathbf{\hat{u}}^{(a)} = \left( \hat{u}_1^{(a)},\dots,\hat{u}_N^{(a)} \right)$ as a \emph{single-node HL solution}, corresponding to the holonomies of a single node $a$, and to $ \left( \mathbf{\hat{u}}^{(1)},\dots,\mathbf{\hat{u}}^{(K)} \right)$ as the \emph{multi-node} HL, corresponding to the collection of all the HL for every nodes of the necklace.
The HL solutions were originally derived for $\mathcal{N}=4$ SYM, corresponding to the limiting case $K=1$ \cite{Benini:2018ywd}. However, these solutions are more general: they solve the BAEs of any $\mathcal{N}=1$ toric gauge theory.
The solution $\{1,N,0\}$ dominates the large $N$ limit, generalizing the $\mathcal{N}=4$ SYM result \cite{Lanir:2019abx,Benini:2020gjh}.

The HL solutions are in one-to-one correspondence with order-$N$ subgroups of $\Z_N\times\Z_N$, namely subsets of $N$ points inside the $N\times N$ lattice. The Weyl group of $\SU(N)$ acts freely by permuting the holonomies and therefore generates $N!$ equivalent representatives of the same solution. Since the SCI is Weyl invariant, all these representatives give the same contribution. Furthermore, since the BAEs depend only on differences of holonomies,  
different choices of the common shift $\bar{u}$ may correspond to inequivalent representatives of the same HL solution, while still yielding the same contribution to the SCI. Although there are $N^2$ possible choices of $\bar{u}$, only $N$ are inequivalent.
Altogether, each HL solution contributes with multiplicity $N\cdot N!$.

While this additional degeneracy only affects the counting of solutions in the simple  case of $\mathcal{N}=4$ SYM, it will play a central role in the analysis of the theories considered here. For this reason we will perform a detailed analysis regarding the structure of the possible choices of $\bar{u}$.
Indeed, the same counting admits a simple lattice interpretation. The action of the Weyl group corresponds to an automorphism of the order-$N$ subgroup associated with $\{m,n,r\}$, while the $N$ inequivalent choices of $\bar{u}$ are in one-to-one correspondence with the $N$ cosets of this subgroup in $\Z_N\times\Z_N$. 
To make the lattice structure of the HL solutions more explicit, we introduce a basis using the cycles of the torus $e_1=(1,0)$ and $e_2=(0,\tau)$.
In this basis, the HL solution can be rewritten as
\begin{equation}
\label{lattice}
\hat{u}=\bar{u}+v_1\hat{\jmath}+v_2\hat{\kappa},
\end{equation}
where we have introduced the vectors
\begin{equation}
    v_1=\left(\frac{1}{m},0\right), \quad     v_2=\left(\frac{r}{N},\frac{1}{n}\right).
\end{equation}
With this notation, each HL solution defines a finite lattice with structure $\mathbb{Z}_m\times\mathbb{Z}_n$. More formally, it is generated by $\mathbb{Z}_m v_1+\mathbb{Z}_n v_2$. Notice that the role of $r$ is only that of rotating the $e_2$ axis, we can therefore obtain a $\mathbb{Z}_m\times\mathbb{Z}_n$ structure, even if we choose $v_2=\left(0,\frac{1}{n}\right)$. This will be more natural in what follows.

\subsubsection*{A new class of solutions}
\label{NewSol}
We now show that when $K>1$ this class of solutions is actually larger. Indeed, the HL solutions \eqref{HLtor} do not make up the full set of discrete solutions of the BAEs \eqref{NKbaesTH1}. Another class of solutions is obtained by playing with the shifts $\bar u$.

\begin{figure}
\centering
\begin{tikzpicture}
\begin{axis}[
    width=11cm,
    height=9cm,
    axis lines=left,
    axis line style={thick},
    xmin=-0.03,
    xmax=1.03,
    ymin=-0.03,
    ymax=1.03,
    xlabel={$\operatorname{Re} u$},
    ylabel={$\operatorname{Im} u$},
    xlabel style={
        at={(axis description cs:1.12,0.04)},
        anchor=north east
    },
    ylabel style={
    rotate=-90,
    at={(axis description cs:0,1.02)},
	anchor=south
	},
    xtick={0,0.25,0.5,0.75,1},
    ytick={0,0.25,0.5,0.75,1},
    xticklabels={$0$,$\frac14$,$\frac12$,$\frac34$,$1$},
    yticklabels={$0$,$\frac14$,$\frac12$,$\frac34$,$1$},
    tick align=outside,
    tick style={thick},
    grid=major,
    grid style={gray!25},
    legend style={
        at={(1.05,0.5)},
        anchor=west,
        draw=none,
        fill=none,
        font=\small,
        row sep=5pt
    },
    legend cell align={left},
    clip=false
]
% Prima famiglia
\addplot[
    only marks,
    mark=*,
    mark size=4pt,
    Petrolio
]
coordinates {
    (0.25,0.25)
    (0.25,0.75)
    (0.75,0.25)
    (0.75,0.75)
};
\addlegendentry{
    \,$\bar{u}$
}
% Seconda famiglia
\addplot[
    only marks,
    mark=square*,
    mark size=4pt,
    Senape
]
coordinates {
    (0,0.25)
    (0,0.75)
    (0.5,0.25)
    (0.5,0.75)
};
\addlegendentry{
    \,$\bar{u}+\frac{1}{4}$
}
% Terza famiglia
\addplot[
    only marks,
    mark=diamond*,
    mark size=5pt,
    Prugna
]
coordinates {
    (0.25,0)
    (0.25,0.5)
    (0.75,0)
    (0.75,0.5)
};
\addlegendentry{
    \,$\bar{u}+\frac{\tau}{4}$
}
% Quarta famiglia
\addplot[
    only marks,
    mark=triangle*,
    mark size=5pt,
    Mattone
]
coordinates {
    (0,0)
    (0,0.5)
    (0.5,0)
    (0.5,0.5)
};
\addlegendentry{
    \,$\bar{u}+\frac{1+\tau}{4}$
}
\end{axis}
\end{tikzpicture}
\caption{We consider the HL solution $\{2,2,0\}$ of $\SU(4)$. Each solution has four holonomies $(u_1,u_2,u_3,u_4)$ that are plotted in the complex plane, with $\tau=\mi$. For each solution the constant term is shifted according to \eqref{HLshift}.
These four choices of $\bar{u}$ complete a lattice in the fundamental cell of the complex torus.}
\label{SU4shiftlattice}
\end{figure}
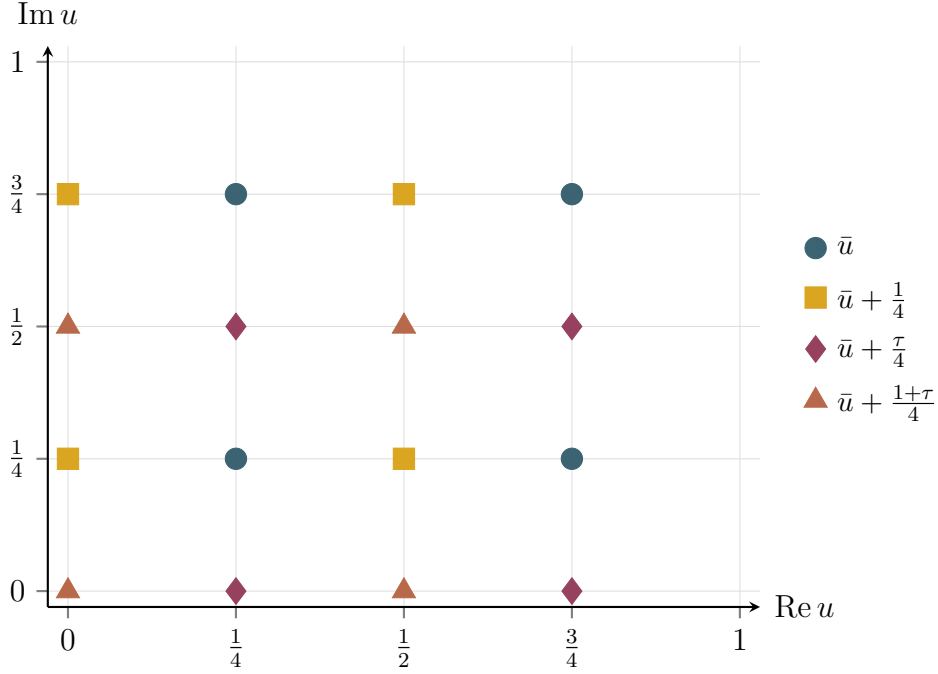

In $K>1$, we actually have the freedom of choosing independent $\bar u$ for each individual node. As mentioned above, for each node $\bar u$ can indeed be chosen in $N=mn$ different ways, while still preserving the constraint \eqref{suconstrTOR}. Without loss of generality, we can fix a reference $\bar u$ and then the precise set of all possible choices for a given $\{m,n,r\}$ solution is determined by the following shifts
\begin{equation}
\label{HLshift}
    \bar{u} \; \longmapsto \; \bar{u}+\frac{\alpha + \beta \tau}{N}\,, 
	\qquad
	\alpha = 0, \dots, n-1\,,
	\quad
	\beta = 0, \dots, m-1\,.
\end{equation}
The constraint \eqref{suconstrTOR} is preserved even if we take $\alpha, \beta=0,\dots,N-1$, but such an extension does not generate new points on the lattice. The shifts \eqref{HLshift} define the $N$ inequivalent embeddings of the HL sublattices into the $\mathbb{Z}_N\times\mathbb{Z}_N$ lattice. An example of this is shown in Figure \ref{SU4shiftlattice}.

In the case of $K=1$, \emph{i.e.} $\mathcal{N} = 4$ SYM, shifting the $N$ holonomies by the same quantity \eqref{HLshift} only gives an overall $N$ degeneracy \cite{Benini:2021ano,Benini:2018mlo}, because the holonomies only enter the SCI as differences, simplifying $\bar u$. 
The same happens for $K>1$ if we shift by the same quantity the fixed terms $\bar{u}$ for all the gauge nodes.
On the other hand, when we take $\bar u=\bar u^{(a)}$, the constant term no longer simplifies in the SCI. In this way, we can indeed construct new candidate solutions to the BAEs and in the following, we prove that these indeed satisfy the BAEs.
Starting from any multi-node HL solution, a new candidate solution is obtained by independently applying a rigid shift to the holonomies of each gauge node:
\begin{equation}
	\label{NKshift}
	\mathbf{\hat{u}}^{(a)} \; \longmapsto \;
	\mathbf{\hat{u}}^{(a)} +\mathbf{s}^{(a)},
\end{equation}
where the shift $\mathbf{s}^{(a)} = \left(s^{(a)},\dots,s^{(a)}\right)$ is defined from \eqref{HLshift} as
\begin{equation}
	\label{HLshift2}
	s^{(a)} = \frac{\alpha^{(a)}+\beta^{(a)}\tau}{N}\,,
	\qquad
	\alpha^{(a)} = 0, \dots, n-1\,,
	\quad
	\beta^{(a)} = 0, \dots, m-1\,.
\end{equation}
 We also define the vector containg all the shift for each node as $s=\left(\mathbf{s}^{(1)}, \dots \mathbf{s}^{(K)}\right)$. 
In what follows, we will refer to the multi-node solution in which each node is left unshifted as the \emph{progenitor}, and to the other multi-node solutions obtained from it by shifting any node as its \emph{descendants}. We will indicate the descentant of a $\{m,n,r\}$ progenitor as 
\begin{equation}
	\label{HLgen}
	\left\{m,n,r\right\}_s\,: \quad
\hat{u}_{i}^{(a)}=\hat{u}_{\hat{\jmath}\hat{\kappa}}^{(a)} = 
	\bar{u}+s^{(a)} +
	\frac{\hat{\jmath}}{m} + 
	\frac{\hat{\kappa}}{n} \left(
		\tau + \frac{r}{m}
	\right), \quad \forall\,a = 1,\dots,K\,.
\end{equation}

All the possible shifts together themselves form a lattice. The lattice can indeed by obtained starting from \eqref{lattice} by switching the roles of the $\mathbb{Z}_m$ and of the $\mathbb{Z}_n$. This procedure does not change the structure of the lattice, but applying this to the HL, we find a new object: 
\begin{equation}
    \mathbb{Z}_n v_1+\mathbb{Z}_m v_2.
\end{equation}
The elements of this lattice act on the HL lattice by shifting the holonomies and creating an inequivalent sublattice to that of the original HL. In this sense, the lattice of the shifts is dual to the HL lattice. 
Notice that in order to respect the $\SU(N)$ constraint \eqref{suconstrTOR} on the holonomies we need to consider $v_2=(0,\frac{1}{n})$.
If we express the elements of $ \mathbb{Z}_n$ and $ \mathbb{Z}_m$ as above using $\alpha$ and $\beta$, we see that this object is precisely the shift of \eqref{HLshift} and that the elements of this lattice are the $s$ defined above.

Lastly, we prove by direct evaluation that this procedure always generates a solution of the BAEs \eqref{NKbaesTH1}, and that the resulting solution is inequivalent to the original one whenever the shifts $s^{(a)}$ are not all equal.
To this end, we first need to rewrite the BAEs using the HL language, where the index $j$ decomposes into the two indices $\hat{\jmath}$ and $\hat{\kappa}$ (and similarly $i$ decomposes into $\hat{\jmath}_0$ and $\hat{\kappa}_0$ in our notation), thus
\begin{equation}
\label{notation}
	\prod_{j=1}^N \; \longrightarrow \; 
	\prod_{\hat{\jmath}=0}^{m-1}\prod_{\hat{\kappa}=0}^{n-1}
	\qquad \text{and} \qquad
	\hat{u}_{ji}^{(a,b)} = \hat{u}_j^{(a)} - \hat{u}_i^{(b)}\; \longrightarrow \;
	\hat{u}_{
		\hat{\jmath} \hat{\jmath}_0
		\hat{\kappa} \hat{\kappa}_0
	}^{(a,b)}=
	\hat{u}_{\hat{\jmath} \hat{\kappa}}^{(a)}-
	\hat{u}_{\hat{\jmath}_0 \hat{\kappa}_0}^{(b)}\,.
\end{equation}
Each $u_{ji}^{(a,b)}$, evaluated on the descendant solution \eqref{NKshift}, takes the form
\begin{equation}
    \label{andreaculo}
	\hat{u}_{
		\hat{\jmath} \hat{\jmath}_0
		\hat{\kappa} \hat{\kappa}_0
	}^{(a,b)} = 
	\frac{(\hat{\jmath} - \hat{\jmath}_0) +
            \tilde{\tau} (\hat{\kappa} - \hat{\kappa}_0)}{m}
	+ \left(s^{(a)}-s^{(b)}\right)\,.
 \end{equation}
We apply to all product of $\theta_1$ functions appearing in the BAEs \eqref{NKbaesTH1} the following identity, introduced in \cite{Hong:2018viz}:
\begin{equation}
    \label{hlshift}
    \prod_{\hat{\jmath}=0}^{m-1}\prod_{\hat{\kappa}=0}^{n-1}
    \theta_1\left(
        \delta \pm 
        \frac{(\hat{\jmath} - \hat{\jmath}_0) +
            \tilde{\tau} (\hat{\kappa} - \hat{\kappa}_0)}{m};\,\tau
    \right) = 
    \zeta
    \prod_{\hat{\jmath}=0}^{m-1}\prod_{\hat{\kappa}=0}^{n-1}
    \theta_1\left(
        \delta \pm 
        \frac{\hat{\jmath} + \tilde{\tau} \hat{\kappa}}{m}
		; \tau
    \right),
\end{equation}
where
\begin{equation}
    \label{coeffc}
    \zeta = 
    (-1)^{n \hat{\jmath}_0 + (m+r)\hat{\kappa}_0}
    \ee^{-\pi \mi m \hat{\kappa}_0 \tau}
    \ee^{\pm \pi \mi \hat{\kappa}_0
    \left[ 
        2 m \delta \pm \tilde{\tau} (2n-1-\hat{\kappa}_0)
        \right]}\,.
\end{equation}
This identity holds for any $\delta$, introduced to absorb all terms that do not depend on the indices labeling the holonomies. 
Each product of $\theta_1$ functions in \eqref{NKbaesTH1} can be recast as the l.h.s. of \eqref{hlshift} by reabsorbing the extra shifts of \eqref{andreaculo}, together with the chemical potentials, into $\delta$.
If we then apply the identity, the terms depending on either $\hat{\jmath}_0$ or $\hat{\kappa}_0$ cancel out.
The Lagrange multiplier $\lambda^{(a)}$ can then be chosen appropriately such that the BAEs \eqref{NKbaesTH1} are satisfied.

Notice that this procedure generalizes the construction of new solutions to the BAEs introduced in \cite{Amariti:2025vjd} for the conifold and SPP singularities with gauge $\SU(2)$ nodes.
We also stress that this proof relies on the assumption that, for each bifundamental chiral $X_{a,b}$, there exists a corresponding bifundamental chiral $X_{b,a}$. This assumption is not satisfied in chiral gauge theories. Indeed, we have explicitly checked that these families of solutions do not solve the BAEs of $\mathrm{F}_0$, $\mathrm{dP}_0$, $\mathrm{dP}_1$.

\section{Aspects of the new Bethe vacua}
\label{newsolprop}
In this section we discuss some properties of the new families of descendant solutions found in Section \ref{NewSol}. We show how our class of solutions generalizes the features of HL solutions to necklace models. We will discuss their numerology, and their transformation properties under modular transformations. We will show that progenitors and descendants sit in orbits of the mapping class group of a torus with $K$ punctures, discussed in Section \ref{Sduality}, identified as the underlying S-duality group of such theories. Lastly, we evaluate their contribution to the SCI, both at finite rank and at large $N$. 

\subsection{Counting of discrete Bethe vacua}
\label{GHLcounting}
We recall that the HL solutions are isomorphic to order-$N$ subgroups $\Z_m \times \Z_n$, with $mn = N$, of $\mathbb{Z}_N\times \mathbb{Z}_N$. They are classified by the possible divisors $d$ of $N$. Accordingly, the total number of HL solutions is given by the divisor function \cite{Benini:2018ywd}:
\begin{equation}
\sigma_1(N)=\sum_{d \,|\, N} d\,.
\end{equation}
As a consequence, the solutions \eqref{HLgen} are divided into $\sigma_1(N)$ families. Additionally, for $\mathcal{N}=4$ SYM there is a discrete $\Z_N$ symmetry associated with the $N$ different choices of $\bar{u}$, as discussed above.
This symmetry also holds for the necklace theory with $K>1$. It corresponds to shifting the gauge holonomies of all gauge nodes by the same quantity,
\begin{equation}
	s^{(a)}=t\,, \quad  \forall \; a=1,\dots,K\,.
\end{equation}
On the other hand, as discussed above, allowing for different shifts for different gauge nodes we obtain new descendant solutions.
Such solutions can be organized according to partitions of $K$ of order $N$.
Consider $k=\left[k_0,\dots,k_{N-1}\right]$ such that 
\begin{equation}
	\sum_{i=0}^{N-1}k_i = K\,,
\end{equation}
where $k_i$ counts the number of nodes with a given shift, i.e.
\begin{equation}
    k_i=\#\left\{\,a \; : \; s^{(a)}=i\,\right\}. 
\end{equation}
Then, the total number of descendants in each family is simply obtained by summing over all possible partitions with the appropriate combinatorial multiplicity. Notice that we must also identify configurations related by the unbroken diagonal $\Z_N$ center symmetry discussed above to obtain the physically distinct solutions. Equivalently, the total number of descendants is obtained by dividing the naive counting $N^K$, as there are $N$ distinct possible shifts for each node, by $N$.
The total number of solution, when taken in account all partitions and HL progenitors, is then
\begin{equation}
	\label{countingGHLeq}
	\#(\text{HL + new solutions}) = N^{K-1} \sigma_1(N)\,.
\end{equation}

\subsubsection*{An example}

Consider $N=3$, $K=3$ and choose for example the $\{1,3,0\}$ HL solution for each single node.
On this solution the shift \eqref{HLshift2} takes the form
\begin{equation}
	\mathbf{s}^{(a)} = \left(\frac{\alpha^{(a)}}{3},\frac{\alpha^{(a)}}{3},\frac{\alpha^{(a)}}{3}\right), \qquad \alpha^{(a)}=0,1,2\,, \quad a=1,2,3\,.
\end{equation}
We introduce a short notation where we indicate the single-node HL progenitor plus each possible shift with a bold letter. In the case at hand, we have 3 options:
\begin{equation}
	\mathbf{a} = \{3,1,0\}\,, \quad
	\mathbf{b} = \{3,1,0\}+
		\biggl(\frac{1}{3},\frac{1}{3},\frac{1}{3}\biggr), \quad
	\mathbf{c} = \{3,1,0\} +
		\biggl(\frac{2}{3},\frac{2}{3},\frac{2}{3}\biggr).
\end{equation}
In this language, the unbroken diagonal center $\Z_3$ acts simultaneously on the holonomies of each node as
\begin{equation}
	\Z_3 : \mathbf{a} \mapsto \mathbf{b} \mapsto \mathbf{c} \mapsto \mathbf{a}\,.
\end{equation}
In the table below, we summarize all the multi-node HL, organized according to their corresponding partitions $\left[k_0,k_1,k_2\right]$ of $K=3$.
\\
\begin{equation}
\begin{array}{cccc}
\toprule
\mu \;\;& \;\;[3,0,0]\;\; & \;\;[2,1,0] \;\;&\;\; [1,1,1]\;\; \\
\midrule
& (\mathbf{a},\mathbf{a},\mathbf{a}) & (\mathbf{a},\mathbf{a},\mathbf{b}) & (\mathbf{a},\mathbf{b},\mathbf{c}) \\
&         & (\mathbf{a},\mathbf{b},\mathbf{a}) & (\mathbf{b},\mathbf{a},\mathbf{c}) \\
&         & (\mathbf{b},\mathbf{a},\mathbf{a}) &         \\
&         & (\mathbf{a},\mathbf{b},\mathbf{b}) &         \\
&         & (\mathbf{b},\mathbf{a},\mathbf{b}) &         \\
&         & (\mathbf{a},\mathbf{a},\mathbf{c}) &         \\
\midrule
& 1       & 6       & 2       \\
\bottomrule
\end{array}
\end{equation}
\\
Counting all these solutions gives the expected number of descendants associated with the HL progenitor $(\mathbf{a},\mathbf{a},\mathbf{a})$:
\begin{equation}
	N^{K-1} = 9 = 1+6+2\,.
\end{equation}
In the table, we have included only one representative for each equivalence class under the action of the diagonal $\Z_3$:
\begin{equation}
	(\mathbf{a},\mathbf{a},\mathbf{a}) \overset{\Z_3}{\mapsto}
	(\mathbf{b},\mathbf{b},\mathbf{b}) \overset{\Z_3}{\mapsto}
	(\mathbf{c},\mathbf{c},\mathbf{c})\,, \quad
	(\mathbf{a},\mathbf{a},\mathbf{b}) \overset{\Z_3}{\mapsto}
	(\mathbf{b},\mathbf{b},\mathbf{c}) \overset{\Z_3}{\mapsto}
	(\mathbf{c},\mathbf{c},\mathbf{a})\,, \quad \dots
\end{equation}
since elements belonging to the same equivalence class correspond to the same solution for the evaluation of the SCI.

\subsection{Action of the S-duality group}
\label{M1K}
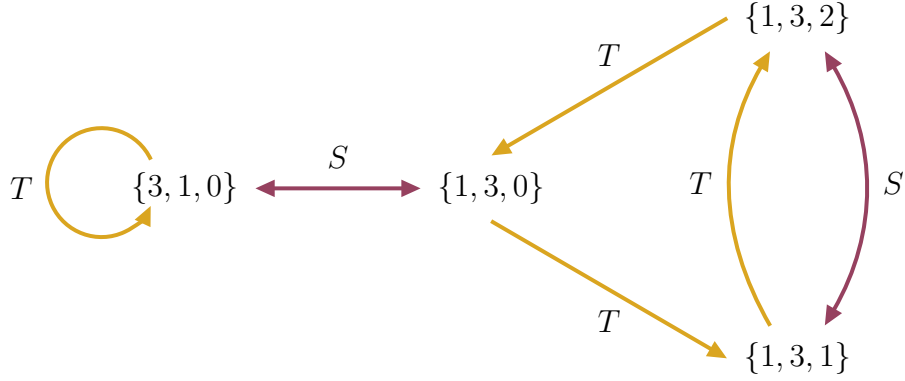
\begin{figure}
\centering
\begin{tikzpicture}[
	scale=0.9,
    >={Triangle[angle=45:8pt]},
    node distance=2.8cm and 3.8cm,
    every node/.style={inner sep=1pt},
    lab/.style={midway, above=6pt, font=\itshape},
    arrlight/.style={->, line width=1.5pt, draw=Senape},
    arrdark/.style={->, line width=1.5pt, draw=Prugna},
    arrdbl/.style={<->, line width=1.5pt, draw=Prugna},
]

\pgfmathsetmacro{\a}{25}
\pgfmathsetmacro{\x}{1.5}
\pgfmathsetmacro{\b}{\x*sin(\a)}
\pgfmathsetmacro{\c}{0.2}
\pgfmathsetmacro{\y}{2.5}

% Nodi
\node (A) at (-\x,0) {};
\node (SU3) at (0,0) {$\{3,1,0\}$};
\node (Z30) at (3*\x,0) {$\{1,3,0\}$};
\node (Z31) at (6*\x,-2.5) {$\{1,3,1\}$};
\node (Z32) at (6*\x,2.5) {$\{1,3,2\}$};

% T arrows 
\draw[arrlight] ($(A)+(1,\b/1.5)$) arc[start angle=\a,end angle=360-\a,radius=0.8];
\node at (-1.6*\x,0) {$T$};
\draw[arrlight] ($(Z30.south)+(0,-\c)$) -- node[lab,below=6pt] {$T$} ($(Z31.west)+(-\c,0)$);
\draw[arrlight] ($(Z32.west)+(-\c,0)$) -- node[lab] {$T$} ($(Z30.north)+(0,\c)$);
\draw[arrlight] ($(Z31.north)+(-2*\c,\c)$)
    to[out=120,in=-120] node[pos=0.52,left=5pt,font=\itshape] {$T$} ($(Z32.south)+(-2*\c,-\c)$);
% S arrows 
\draw[arrdbl] ($(SU3.east)+(\c,0)$) -- node[lab] {$S$} ($(Z30.west)+(-\c,0)$);
\draw[arrdbl] ($(Z31.north)+(2*\c,\c)$) to[out=60,in=-60] node[pos=0.52,right=5pt,font=\itshape] {$S$} ($(Z32.south)+(2*\c,-\c)$);
\end{tikzpicture}
\caption{HL solutions for $\SU(3)$ organized as orbits of $\mathrm{PSL}(2,\Z)$.}
\label{su3orbits}	
\end{figure}

On the HL solutions there is a natural action of $\mathrm{SL}(2,\mathbb{Z})$ inherited by the modular invariance of the BAEs. 
As a consequence of the correspondence of the HL solutions with the lattices, such action acts as the induced action of the $\mathrm{SL}(2,\mathbb{Z})$ of the torus on its sublattices. Explicitly, we have \cite{Benini:2018ywd}
\begin{equation}
	\label{Sdualityhol}
	S: 
	\begin{cases}
		u \mapsto \frac{u}{\tau} \\
		\tau \mapsto -\frac{1}{\tau}
	\end{cases},
	\qquad \qquad
	T: 
	\begin{cases}
		u \mapsto u \\
		\tau \mapsto \tau+1
	\end{cases},
    \qquad \qquad
    C: 
	\begin{cases}
		u \mapsto -u \\
		\tau \mapsto \tau
	\end{cases}.
\end{equation}
under which the HL solutions organize in $\mathrm{PSL}(2,\mathbb{Z})$ orbits.
Under the action of these generators the HL transform as
\begin{equation}
	\label{SdualityHL}
	S:\{m,n,r\} \mapsto
	\left\{\mathrm{gcd}(n,r),\frac{mn}{\mathrm{gcd}(n,r)},\frac{m(n-r)}{\mathrm{gcd}(n,r)}\right\},
	\quad
	T:\{m,n,r\} \mapsto \{m,n,r+m\}\,.
\end{equation}
See Figure \ref{su3orbits} for the explicit example of $\SU(3)$.

There is another $\mathrm{SL}(2,\Z)$ group in $\mathcal{N} =4$ SYM, the S-duality group. This group acts on the vacua of $\mathcal{N}=1^*$ SYM, discussed in the introduction, in precisely the same way as the $\mathrm{SL}(2,\Z)$ symmetry of the BAEs acts on the HL solutions. This fact constitutes a non-trivial check for Conjecture \ref{Conjecture}, and suggests an identification between the two $\mathrm{SL}(2,\Z)$ on the two sides of the correspondence. It is then tempting to ask whether this holds more generally, namely if the S-duality group $\mathcal{M}(1,K)$ of the necklace models acts analogously on the solutions \eqref{HLgen}.
Notice that the modular invariance of the BAEs is a priori unrelated to the S-duality group of the underlying theory, since the superconformal index is independent of the gauge couplings. 
Although the two groups are in general unrelated, we will show that such an identification is indeed justified and well defined on this class of solutions. In particular, for elliptic models the solutions \eqref{HLgen} naturally organize into orbits of the S-duality group, which in these theories is larger than the $\mathrm{SL}(2,\mathbb{Z})$ symmetry of the BAEs. 
As reviewed in Section \ref{Sduality}, the S-duality group  for the $\mathcal{N}=2$ necklace theory is identified with the mapping class group of the torus with $K$ punctures.
Starting from the basis $(S,T,\sigma_1,\omega)$, we define the action for these generators on the gauge holonomies $u^{(a)}$:
\begin{equation}
	\label{M1Kbasis}
	\begin{aligned}
		S\,&:
		\left(\tau,u^{(1)},\dots,u^{(K)}\right) \longmapsto \left(-\tfrac{1}{\tau},\tfrac{u^{(1)}}{\tau},\dots,\tfrac{u^{(K)}}{\tau}\right)\\
		T\,&:
		\left(\tau,u^{(1)},\dots,u^{(K)}\right) \longmapsto \left(\tau+1,\,u^{(1)},\dots,u^{(K)}\right) \\
		\sigma_1\,&:
		\left(\tau,u^{(1)},\dots,u^{(K)}\right) \longmapsto \left(\tau,\,u^{(2)},\,u^{(1)},\dots,u^{(K)}\right) \\
		\omega\,&:
		\left(\tau,u^{(1)},\dots,u^{(K)}\right) \longmapsto \left(\tau,\,u^{(2)},\,u^{(3)},\dots,u^{(K)},\,u^{(1)}+\tfrac{\tau}{N}\right)\,.
	\end{aligned}
\end{equation}
The generators $S$ and $T$ change the HL progenitor as in the $\mathcal{N}=4$ SYM case, while $\sigma_1$ and $\omega$, map one descendant to another.
The same pattern is observed at the level of the global structure: $S$ and $T$ change the lattice, whereas $\sigma_1$ and $\omega$ map the charge lattice into itself. Therefore, our solutions realize orbits of the mapping class group $\mathcal{M}(1,K)$.

\subsubsection*{An example}
Consider once more the case of $N=3$, $K=3$.
We first consider the HL solution $(\mathbf{a},\mathbf{a},\mathbf{a})$, using the notation introduced in Section \ref{GHLcounting}. We then look at the action of the generators $\sigma_1$ and $\omega$ and we depict the orbits generated by them in Figure \ref{M1KorbitsSU3internal}.
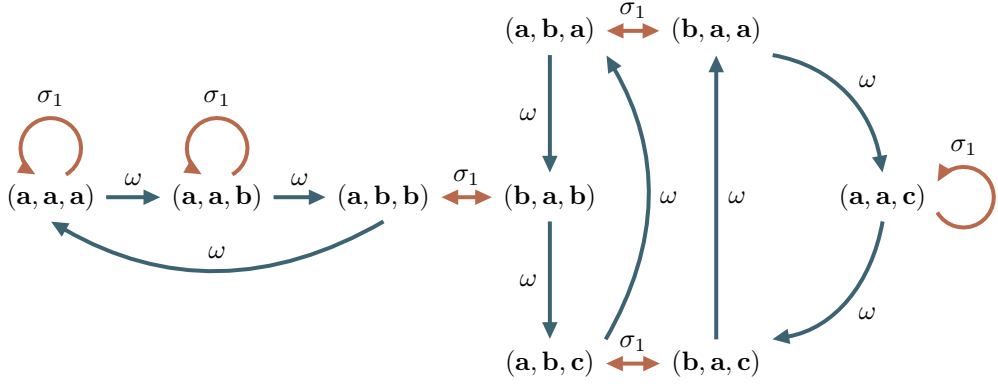
\begin{figure}
\centering
\begin{tikzpicture}[
	scale=1,
    >={Triangle[angle=45:7pt]},
	lab/.style={font=\footnotesize},
    sigma/.style={->, line width=1.5pt, draw=Mattone},
	sigmaD/.style={<->, line width=1.5pt, draw=Mattone},
    omega/.style={->, line width=1.5pt, draw=Petrolio},
]

\pgfmathsetmacro{\x}{2.2}
\pgfmathsetmacro{\a}{30}
\pgfmathsetmacro{\r}{0.4}
\pgfmathsetmacro{\h}{0.1}

% Nodi
\node[lab] (AAA) at (0,0) {$(\mathbf{a},\mathbf{a},\mathbf{a})$};
\node[lab] (AAB) at (\x,0) {$(\mathbf{a},\mathbf{a},\mathbf{b})$};
\node[lab] (ABB) at (2*\x,0) {$(\mathbf{a},\mathbf{b},\mathbf{b})$};
\node[lab] (BAB) at (3*\x,0) {$(\mathbf{b},\mathbf{a},\mathbf{b})$};
\node[lab] (ABA) at (3*\x,\x) {$(\mathbf{a},\mathbf{b},\mathbf{a})$};
\node[lab] (ABC) at (3*\x,-\x) {$(\mathbf{a},\mathbf{b},\mathbf{c})$};
\node[lab] (BAA) at (4*\x,\x) {$(\mathbf{b},\mathbf{a},\mathbf{a})$};
\node[lab] (BAC) at (4*\x,-\x) {$(\mathbf{b},\mathbf{a},\mathbf{c})$};
\node[lab] (AAC) at (5*\x,0) {$(\mathbf{a},\mathbf{a},\mathbf{c})$};

% T arrows 
\draw[omega] (AAA) -- node[lab, above] {$\omega$} (AAB);
\draw[omega] (AAB) -- node[lab, above] {$\omega$} (ABB);
\draw[omega] (ABA) -- node[lab, left] {$\omega$} (BAB);
\draw[omega] (BAB) -- node[lab, left] {$\omega$} (ABC);
\draw[omega] (BAC) -- node[lab, right] {$\omega$} (BAA);
\draw[omega] (ABC.north east) to[out=60,in=-60] node[lab, right] {$\omega$} (ABA.south east);
\draw[omega] (BAA.south east) to[out=-10,in=100] node[lab, right, yshift=5pt] {$\omega$} (AAC.north);
\draw[omega] (AAC.south) to[out=-100,in=10] node[lab, right, yshift=-5pt] {$\omega$} (BAC.north east);
\draw[omega] (ABB.south) to[out=210,in=-30] node[lab, above] {$\omega$} (AAA.south);

\draw[sigmaD] (ABB) -- node[lab, above] {$\sigma_1$} (BAB);
\draw[sigmaD] (ABA) -- node[lab, above] {$\sigma_1$} (BAA);
\draw[sigmaD] (ABC) -- node[lab, above] {$\sigma_1$} (BAC);

\draw[sigma] ($(AAA)+(\r/2,\r-\h)$)
    arc[start angle=-90+\a,end angle=270-\a,radius=\r]
    node[lab, midway, above] {$\sigma_1$};

\draw[sigma] ($(AAB)+(\r/2,\r-\h)$)
    arc[start angle=-90+\a,end angle=270-\a,radius=\r]
    node[lab, midway, above] {$\sigma_1$};

\draw[sigma] ($(AAC.east)+(0,-\r/2)$)
    arc[start angle=-180+\a,end angle=180-\a,radius=\r]
    node[lab, xshift=10pt, yshift=13pt] {$\sigma_1$};

\end{tikzpicture}
\caption{Closed orbits of the generators $\sigma_1$ and $\omega$ of the mapping class group $\mathcal{M}(1,K)$ acting on the family of solutions generated by a given HL progenitor.}
\label{M1KorbitsSU3internal}
\end{figure}
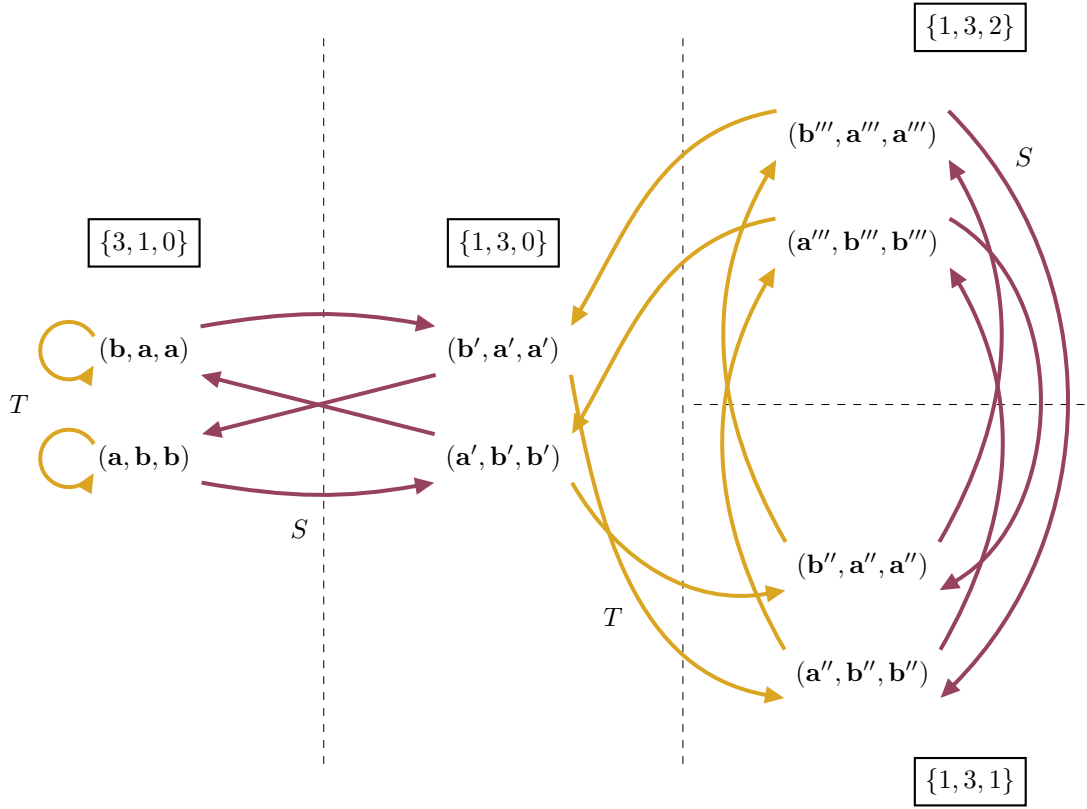
\begin{figure}
\centering
\begin{tikzpicture}[
	scale=0.95,
    >={Triangle[angle=45:8pt]},
	lab/.style={font=\footnotesize},
    Tarrow/.style={->, line width=1.5pt, draw=Senape},
    Sarrow/.style={->, line width=1.5pt, draw=Prugna},
	dashing/.style={-, line width=0.3pt, dashed},
	box/.style={rectangle, draw, thick},
]

\pgfmathsetmacro{\x}{1.5}
\pgfmathsetmacro{\y}{5}
\pgfmathsetmacro{\z}{1.5}
\pgfmathsetmacro{\a}{30}
\pgfmathsetmacro{\r}{0.4}
\pgfmathsetmacro{\h}{0.3}

\node[lab,box] (HL1) at (0,\x+\z) {$\{3,1,0\}$};
\node[lab] (BAA1) at (0,\x) {$(\mathbf{b},\mathbf{a},\mathbf{a})$};
\node[lab] (ABB1) at (0,0) {$(\mathbf{a},\mathbf{b},\mathbf{b})$};

\node[lab,box] (HL2) at (\y,\x+\z) {$\{1,3,0\}$};
\node[lab] (BAA2) at (\y,\x) {$(\mathbf{b}',\mathbf{a}',\mathbf{a}')$};
\node[lab] (ABB2) at (\y,0) {$(\mathbf{a}',\mathbf{b}',\mathbf{b}')$};

\node[lab,box] (HL3) at (2*\y+\z,-2*\x-\z) {$\{1,3,1\}$};
\node[lab] (BAA3) at (2*\y,-\x) {$(\mathbf{b}'',\mathbf{a}'',\mathbf{a}'')$};
\node[lab] (ABB3) at (2*\y,-2*\x) {$(\mathbf{a}'',\mathbf{b}'',\mathbf{b}'')$};

\node[lab,box] (HL4) at (2*\y+\z,3*\x+\z) {$\{1,3,2\}$};
\node[lab] (ABB4) at (2*\y,2*\x) {$(\mathbf{a}''',\mathbf{b}''',\mathbf{b}''')$};
\node[lab] (BAA4) at (2*\y,3*\x) {$(\mathbf{b}''',\mathbf{a}''',\mathbf{a}''')$};

\node (d11) at (\y/2,3*\x+\z) {};
\node (d12) at (\y/2,-2*\x-\z) {};
\draw[dashing] (d11) -- (d12);

\node (d21) at (3*\y/2,3*\x+\z) {};
\node (d22) at (3*\y/2,-2*\x-\z) {};
\draw[dashing] (d21) -- (d22);

\node (d31) at (3*\y/2,\x/2) {};
\node (d32) at (2*\y+3*\z/2+\z/2+0.3,\x/2) {};
\draw[dashing] (d31) -- (d32);

%\draw[omega] (AAA) -- node[lab, above] {$\omega$} (AAB);

\draw[Sarrow] (ABB1.south east) to[out=-10,in=190]
    node[lab, below, xshift=-7pt,yshift=-5pt] {$S$}
    (ABB2.south west);

%\draw[Sarrow] (ABB2.north west) to[out=180,in=-30]
%    node[lab, below, xshift=8pt] {} (BAA1.south east);

\draw[Sarrow] (ABB2.north west) -- (BAA1.south east);

\draw[Sarrow] (BAA1.north east) to[out=10,in=170]
    node[lab, below, xshift=-7pt] {}
    (BAA2.north west);

%\draw[Sarrow] (BAA2.south west) to[out=180,in=30]
%    node[lab, above, xshift=8pt] {} (ABB1.north east);

\draw[Sarrow] (BAA2.south west) -- (ABB1.north east);

\draw[Tarrow] ($(BAA1)+(-\r-\h,\r/2)$)
    arc[start angle=\a,end angle=360-\a,radius=\r]
    node[lab, midway, xshift=-8pt,yshift=-20pt] {$T$};

\draw[Tarrow] ($(ABB1)+(-\r-\h,\r/2)$)
    arc[start angle=\a,end angle=360-\a,radius=\r]
    node[lab, midway, xshift=-8pt] {};

\draw[Tarrow] (ABB2.south east) to[out=-60,in=190]
    node[lab, below, xshift=-7pt,yshift=-5pt] {}
    (BAA3.south west);

\draw[Tarrow] (BAA2.south east) to[out=-80,in=170]
    node[lab, below, xshift=-7pt,yshift=-5pt] {$T$}
    (ABB3.south west);

\draw[Tarrow] (ABB3.north west) to[out=120,in=-120]
    node[lab, below, xshift=-7pt,yshift=-5pt] {}
    (ABB4.south west);

\draw[Tarrow] (BAA3.north west) to[out=120,in=-120]
    node[lab, below, xshift=-7pt,yshift=-5pt] {}
    (BAA4.south west);

\draw[Tarrow] (ABB4.north west) to[out=190,in=60]
    node[lab, below, xshift=-7pt,yshift=-5pt] {}
    (ABB2.north east);

\draw[Tarrow] (BAA4.north west) to[out=190,in=60]
    node[lab, below, xshift=-7pt,yshift=-5pt] {}
    (BAA2.north east);

\draw[Sarrow] (ABB3.north east) to[out=60,in=-60]
    node[lab, below, xshift=-7pt,yshift=-5pt] {}
    (ABB4.south east);

\draw[Sarrow] (BAA3.north east) to[out=60,in=-60]
    node[lab, below, xshift=-7pt,yshift=-5pt] {}
    (BAA4.south east);

\draw[Sarrow] (ABB4.north east) to[out=-30,in=30]
    node[lab, below, xshift=-7pt,yshift=-5pt] {}
    (BAA3.south east);

\draw[Sarrow] (BAA4.north east) to[out=-45,in=45]
    node[lab, below, xshift=-7pt,yshift=-5pt] {}
    (ABB3.south east);

\node[lab] (S) at (2*\y+3*\z/2,3*\x-\h) {$S$};

\end{tikzpicture}
\caption{Closed orbits of the two generators $S$ and $T$ of $\mathcal{M}(1,K)$ acting on the representatives $(\mathbf{a},\mathbf{b},\mathbf{b})$ and $(\mathbf{b},\mathbf{a},\mathbf{a})$ of each family of solutions.}
\label{M1KorbitsSU3external}	
\end{figure}
A direct application of $\omega$ does not necessarily generate all the elements shown in the figure. 
However, we can recover them by combining $\omega$ with the action of the central $\Z_3$ symmetry:
\begin{equation}
(\mathbf{a},\mathbf{b},\mathbf{c}) \overset{\omega}{\longmapsto}
(\mathbf{b},\mathbf{c},\mathbf{b}) \overset{\Z_3}{\longmapsto}
(\mathbf{c},\mathbf{a},\mathbf{c}) \overset{\Z_3}{\longmapsto}	(\mathbf{a},\mathbf{b},\mathbf{a}) \,.
\end{equation}

Now we consider, for each HL progenitor $(\mathbf{a},\mathbf{a},\mathbf{a})$, two descendants: $(\mathbf{b},\mathbf{a},\mathbf{a})$ and $(\mathbf{a},\mathbf{b},\mathbf{b})$. We then consider the action of the generators $S$ and $T$. These generators map the descendants of an HL progenitor into other descendants of another HL progenitor. A descendant is not necessarily trivially mapped into the corresponding descendant of another family, but nevertheless the orbits that are generated are closed, as shown in Figure  \ref{M1KorbitsSU3external}. 

The choice of $N=3$ as an example is motivated by the fact that it already encodes all the relevant features, while still allowing for a simple graphical representation.
For $N>3$ a graphical realization is still possible, but it becomes considerably more involved, since the shifts \eqref{HLshift2} require linear combinations of the basis elements \eqref{M1Kbasis}. Using this example we were instead able to represent separately the closed orbits of $\sigma_1$, $\omega$ and $S$, $T$.

\subsection{Relation with vacua of \texorpdfstring{$\mathcal{N} = 1^*$}{} deformations}
\label{subsec:countingspin}

As reviewed in Section \ref{necklacereview} necklace models admit $\mathcal{N}=1^*$ deformations with a very rich structure of massive and massless vacua, similarly to the case of $\mathcal{N} = 4$ SYM. 
In light of the Conjecture \ref{Conjecture}, we can therefore compare our solutions \eqref{HLgen} of the BAEs to such vacua and extend the conjectural correspondence. We preliminarily notice that the total number of solutions 
\begin{equation}
    N^{K-1} \sigma_1(N)
\end{equation}
agrees with the number of massive vacua.

We propose a correspondence between our class of solutions and the massive sector of vacua of $\mathcal{N}=1^*$ deformations. Specifically:
\begin{itemize}
    \item The partition 
    \begin{equation*}
        \{\mathcal{A}_1,...,\mathcal{A}_1, \mathcal{A}_2,...,\mathcal{A}_2,......,\mathcal{A}_K,...,\mathcal{A}_K\},
    \end{equation*}
    with $n$ equal subsets $\mathcal{A}_i$ of $m$ elements for each $i=1,...,K$ is associated with the whole family of descendants $\{m,n,r\}_s$ of the progenitor $\{m,n,r\}$. Notice that the number of vacua in such partition is $N^{K-1} n$ associated to the low-energy confinement of the unbroken $\SU(n)^K$ SYM theory. 
    \item For fixed $m,n$ each class of $N^{K-1}$ descendants is labeled by a different value of $r\in \mathbb{Z}_n$. They are mapped into the corresponding families of $N^{K-1}$ confining vacua, and permuted via the action of $T$, as it also happens for the limiting case $K=1$ of $\mathcal{N} = 4$ SYM. The mapping is a one-to-one mapping between vacua and not just families: for fixed $m,n,r$ the $N^{K-1}$ descendants labeled by $s$ are mapped in a one-to-one correspondence with a corresponding vacuum. 
    \item The precise one-to-one correspondence between massive vacua of $\mathcal{N} =1^*$ systems and solutions to the BAEs can be obtained by following the orbits under the duality group $\mathcal{M}(1,K)$ acting on both sides of the correspondence. In fact, this property constitutes a highly non-trivial consistency check for the correspondence to make sense in the first place. Conversely,  the fact that $\mathcal{M}(1,K)$ acts on the set \eqref{HLgen} can also be interpreted backwards as a consequence of a relation between a specific sector of solutions to the BAEs and a corresponding sector of vacua of $\mathcal{N}=1^*$ systems. 
\end{itemize}

While we have not studied possible continuous solutions for these theories, we refrain to formulate any conjecture on their possible relation with massless vacua of such models. Indeed, it seems that the underlying origin behind such a correspondence consists in a relation between the BAEs describing the superconformal index and the equations arising from Calogero-Moser systems describing the vacua of such $\mathcal{N}=1^*$ deformations. In the case of $\mathcal{N} = 4$ $\SU(N)$ SYM the numerology of solutions seem to match beyond the massive sector of vacua. In general, however, these two equations are very different and a precise correspondence of solutions beyond specific sectors is unexpected. Indeed, even for $\mathcal{N} = 4$ $\SU(N)$ SYM a precise formulation of the correspondence can be drawn only modulo extra degeneracies due to Weyl symmetry and center symmetry actions producing an overall $N \cdot N!$ factor on each HL solution. Indeed, the precise correspondence must be formulated modulo these global redundancies by looking at distinct inequivalent contributions to the SCI. Beyond the case of algebra $A_n$ the situation is even more complicated as discussed in \cite{Fazzi:2026wkb} as the BAEs are related to untwisted Calogero-Moser systems which do not describe the vacua of $\mathcal{N}=1^*$ for non-simply laced algebras, so at most only a special subset of solutions could be universal and related to vacua of $\mathcal{N}=1^*$ models. 

Indeed, for the case of necklace models partial results \cite{AMM} on the BAEs of $\mathbb{C}^2/\mathbb{Z}_2\times \mathbb{C}$ seem to suggest a mismatch between other solutions to the BAEs beyond the class \eqref{HLgen} and massless vacua of $\mathcal{N}=1^*$ systems. For this reason we conjecture that only specific sectors of solutions might be universal to both systems.

\subsection{\texorpdfstring{Comments on the finite $N$ evaluation}{}}
\label{finiteN}

We now focus on the evaluation at finite rank $N$ of the contributions of the class of solutions \eqref{HLgen}. The SCI evaluated on such class can be organized into families of $N^{K-1}$ elements according to the distinct phases of the massive deformations of the theory, as discussed in \ref{necklacereview}, by virtue of the aforementioned correspondence. Each solution \eqref{HLgen} is associated with a degeneracy factor $N \cdot N!^K$, where the first factor comes from the equivalence of solutions under the diagonal center $\Z_N$ action, while the latter arises as a degeneracy due to the action of the Weyl group on each node. The second factor cancels with the corresponding term at the denominator of \eqref{SCIneck} and the index can be schematically expanded as
\begin{equation}
\mathcal I(\Delta,\tau)=N\sum_{(m,n,r)}\sum_{\lambda\in\mathcal P(K)}\sum_{\sigma\in\mathfrak S(\lambda)}
I\!\left(\hat u{(m,n,r;\lambda,\sigma)},\Delta;\tau\right),
\end{equation}
with  $m n = N$, $r \in \Z_n$, while $s$ in \eqref{HLgen} parameterized by $\mathcal{P}(K)$ partitions of $K$ of order $N$, which specify the number of nodes with a given shift $s^{(a)}=i$, and $\mathfrak{S}(\lambda)$, the set of all distinct permutations of $(\mathbf{s}^{(1)},...,\mathbf{s}^{(K)})$. According to the BA formula \eqref{baf}, $I(\hat{u},\Delta,\tau) = \mathcal{Z}(\hat{u},\Delta,\tau)\mathcal{H}^{-1}(\hat{u},\Delta,\tau)$.

In special regions of the parameter space the formula can be simplified. For instance, looking at the fixed point with all baryonic fugacities turned off an extra relative degeneracy factor associated to the cardinality of $\mathfrak{S}(\lambda)$ arises, as turning off baryonic fugacities renders all nodes indistinguishable

\begin{equation}
\mathcal I(\Delta;\tau)=N\sum_{(m,n,r)}\sum_{\lambda\in\mathcal P(K)}
\mathfrak{S}(\lambda) \,I\!\left(\hat u^{(m,n,r;\lambda)},\Delta;\tau\right).
\end{equation}

We can explicitly evaluate the index on our class of solutions and compare the result with the $q-$expansion of the index.
We choose the necklace theory with $N=K=2$ as a working example.

In the $\mathcal{N}=1$ formalism, the matter content consists of four bifundamental chirals $X_{12}^{(1,2)}$, $X_{21}^{(1,2)}$, originating from the $\mathcal{N}=2$ hypermultiplets, and two adjoint chirals $\Phi_{1}$, $\Phi_{2}$, originating from the $\mathcal{N}=2$ vector multiplets.
The quiver of the theory is shown in Figure \ref{2necklace2}, while the superpotential in \eqref{Wneck} reduces to 
\begin{equation}
	\label{22sup}
	W = 
	X_{21}^{(1)} \Phi_{1} X_{12}^{(1)} -
	X_{21}^{(2)} \Phi_{1} X_{12}^{(2)} +
	X_{12}^{(1)} \Phi_{2} X_{21}^{(1)} -
	X_{12}^{(2)} \Phi_{2} X_{21}^{(2)} 
	\,.
\end{equation}
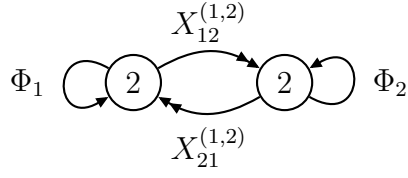
\begin{figure}
	\centering
	\begin{tikzpicture}[
        box/.style={rectangle, draw, thick},
        ]
        \pgfmathsetmacro{\x}{1}
        \pgfmathsetmacro{\y}{1}
		\pgfmathsetmacro{\z}{2.4}
        % Nodes
        \node[circle,draw, thick] (l) at (-\x, 0) {$2$};
        \node[circle,draw, thick] (r) at (\x, 0) {$2$};
        % Bifundamentals
        \node at (0, \y-0.2) {$X_{12}^{(1,2)}$};
        \node at (0, -\y+0.15) {$X_{21}^{(1,2)}$};
        \draw[->>, thick, >={Triangle[angle=45:5pt]}] (l) to[out=30, in=150, looseness=1.1] (r);
        \draw[->>, thick, >={Triangle[angle=45:5pt]}] (r) to[out=-150, in=-30, looseness=1.1] (l);
		% Adjoints
		\node at (\z, 0) {$\Phi_{2}$};
        \node at (-\z, 0) {$\Phi_{1}$};
		\draw[->, thick, >={Triangle[angle=45:5pt]}, looseness=6] (l) to[out=150, in=210] (l);
		\draw[->, thick, >={Triangle[angle=45:5pt]}, looseness=6] (r) to[out=-30, in=30] (r);
	\end{tikzpicture}
	%\vspace{-1em}
	\caption{Quiver diagram of the $\SU(2) \times \SU(2)$ necklace theory in $\mathcal{N}=1$ formalism.}
	\label{2necklace2}
\end{figure}
We parametrize the Cartan subalgebra of the global symmetry group following the prescription proposed in \cite{Butti:2005vn}.
This choice determines the corresponding assignments of chemical potentials, summarized in the table below. \\
{
\setlength{\arraycolsep}{10pt}
\begin{equation}
\label{Dassign}
\begin{array}{cccccc}
\toprule
\Phi_{1} & \Phi_{2} &
X_{12}^{(1)} & X_{12}^{(2)} &
X_{21}^{(1)} & X_{21}^{(2)} \\
\midrule
\Delta_1 & \Delta_1 &
\Delta_2 + \Delta_3 & \Delta_3 + \Delta_4 &
\Delta_4 & \Delta_2 \\
\bottomrule
\end{array}
\end{equation}
}\\
The chemical potentials are subject to the constraint \eqref{Dconstr} that in this basis translates to
\begin{equation}
	\label{deltaconstr}
	\Delta_1 + \Delta_2 + \Delta_3 + \Delta_4 = 2 \tau\,.
\end{equation}
Our set of solutions is listed below, using the notation $\left(u_1^{(1)},u_2^{(1)},u_1^{(2)},u_2^{(2)}\right)$.
{\footnotesize
\begin{equation}
	\label{22sol}
	\begin{aligned}
		\mathfrak{M}_{\text{BAE}}=\Biggl\{
			&\left( \frac{1}{4},\frac{3}{4},\frac{1}{4},\frac{3}{4} \right),
			\left( \frac{\tau}{4},\frac{3\tau}{4},\frac{\tau}{4},\frac{3\tau}{4} \right),
			\left( \frac{1+\tau}{4},\frac{3+3\tau}{4},\frac{1+\tau}{4},\frac{3+3\tau}{4} \right), \\
			&\left( \frac{1+2\tau}{4},\frac{3+2\tau}{4},\frac{1}{4},\frac{3}{4} \right),
			\left( \frac{2+\tau}{4},\frac{2+3\tau}{4},\frac{\tau}{4},\frac{3\tau}{4} \right),
			\left( \frac{3+\tau}{4},\frac{1+3\tau}{4},\frac{1+\tau}{4},\frac{3+3\tau}{4} \right)\,
		\Biggr\}.
	\end{aligned}
\end{equation}}
\!\!The first three elements in the set are the usual HL progenitors, while the last three are the new descendant solutions.

Using the BA formula \eqref{baf}, we evaluate the SCI on this set. The computation follows the approach of \cite{Benini:2021ano,Lezcano:2021qbj,Amariti:2025vjd}.
We choose a suitable parametrization of the flavor fugacities $y_\mu= \ee^{2 \pi \mi \Delta_\mu}$:
\begin{equation}
	q=t^2\,, \quad
	y_1 = a b t\,, \quad
	y_2 = a^{-1} b t\,, \quad
	y_3 = b^{-1} c t\,, \quad
	y_4 = b^{-1} c^{-1} t\,.
\end{equation}
We then perform a small $t$ expansion of \eqref{baf} and compare the result with the corresponding expansion of the matrix integral \eqref{SCIneck} in the same regime
\begin{equation}
	\label{directev}
	\mathcal{I} = 1 + \left(\frac{b^2}{a^2} + 2 a^2 b^2 + \frac{1}{b^2 c^2} + \frac{1}{a c}\right) t^2 + \order{t^4}\,.
\end{equation}
Taking into account that each solution in \eqref{22sol} has multiplicity 8, the small $t$ expansion of \eqref{baf} becomes:
\begin{equation}
	\label{22fail}
	\sum_{\hat{u} \in \mathfrak{M}_{\text{BAE}}}
	\mathcal{Z}\left(\hat{u},\Delta;\tau\right)
	\mathcal{H}\left(\hat{u},\Delta;\tau\right)^{-1}
	= \sum_{n=-2}^{3} \mathcal{I}_n \, t^{n} + \order{t^4}\,,
\end{equation}
where each coefficient $\mathcal{I}_n$ is a function of the fugacities $a$, $b$, $c$.

This expansion does not coincide with the direct evaluation \eqref{directev}. Moreover, it has divergent terms, also referred to as \emph{tachyonic}, signaling that the contributions of other solutions are needed.

In \cite{Amariti:2025vjd}, two examples of evaluation of the SCI were presented.
In the case of the $\SU(2)^2$ conifold theory, the set \eqref{22sol} was sufficient to reproduce the SCI for any fugacities.
In the case of $\SU(2)^3$ suspended pinch point theory (SPP), the analogous generalization of \eqref{22sol} to $K=3$ reproduces the SCI in a specific subregion of fugacities.
Here, instead, we cannot even identify a subregion of fugacities in which such divergences cancel.
The expressions of the coefficients $\mathcal{I}_n$ of \eqref{22fail} for generic fugacities $a$, $b$, $c$ are not particularly insightful. Therefore, we only display the coefficient associated to the divergent terms, as well as those in the special case $a=b=c$, where their expressions take a simpler form:

{\footnotesize
\begin{equation}
	\begin{aligned}
	&\mathcal{I}_{-2} &&\!\!\!\!\!\!= 
	\frac{
		5 a^{22} \!-\!
		14 a^{20} \!-\!
		326 a^{18} \!-\! 
		1072 a^{16} \!-\!
		654 a^{14} \!+\!
		761 a^{12} \!-\! 
		1530 a^{10} \!-\!
		1645 a^8 \!-\!
		579 a^6 \!-\!
		27 a^4 \!-\!
		20 a^2 \!-\! 
		19
	}{
		\left(a^2-1\right)^2
		\left(a^2+1\right)
		\left(a^4+6 a^2+1\right)^4
	}
	\\
	&
	\\
	&\mathcal{I}_{-1} &&\!\!\!\!\!\!=
	\frac{
		a^2
		\left(
			3 a^{14}
			+24 a^{12}
			+109 a^{10}
			+244 a^8
			+21 a^6
			-80 a^4
			-53 a^2
			-12
		\right)
	}{
		\left(a^2-1\right)^2
		\left(a^2+1\right)
		\left(a^4+6 a^2+1\right)^3
	}\,.
	\end{aligned}
\end{equation}
}

We conclude the analysis with a speculative argument on the nature of the divergence of the SCI for this case. 
It is indeed consistent to expect  a divergent result if the correspondence between the solutions of the BAEs and the vacua of the integrable system holds beyond the massive case.
Indeed, for the rank two $\mathbb{C}^3/\mathbb{Z}_2$ model deformed to $\mathcal{N}=1$  through massive adjoints,
the set of vacua found in \cite{Hollowood:2002zk} consists of six massive vacua, corresponding to the six Bethe vacua found here, and in addition two extra massless vacua.
Then, according to the Conjecture \ref{Conjecture}, 
we expect that such  two extra massless vacua  correspond to two continuous solutions of the BAEs. 
In presence of continuous solutions the index is not guaranteed to be fully reproduced by the discrete solutions to the BAEs, as observed for example in the case of $\mathcal{N}=4$ $\SU(3)$ SYM  in \cite{Benini:2021ano}.
Therefore, in order to compute the SCI one needs to identify the continuous solutions and evaluate their contributions\footnote{Actually we cannot exclude in this case the existence of further discrete solutions. However, we have scanned over a large range of rational values for the holonomies 
without finding any numerical evidence for the existence of such solutions.}.
A general prescription to account for such continuous solutions was proposed \cite{Cabo-Bizet:2024kfe}, where indeed it was shown that they cancel the divergent contributions found  in \cite{Benini:2021ano} .
Notice also that similar divergencies do not arise in the evaluations of the SCI for the conifold at rank two 
 \cite{Amariti:2025vjd}, where the discrete Bethe vacua are sufficient to compute the index.
The interesting aspect of this fact is that the conifold theory can be obtained from the $\C^3/\Z_2$ orbifold by introducing mass deformations for the adjoint matter fields with $m_{\Phi_1}=-m_{\Phi_2}$. This is compatible with the interpretation that suitably chosen mass deformations ``lift''  the massless vacua, removing the need for contributions from continuous solutions. We leave a more detailed analysis on the continuous solutions for the BAEs studied here to future investigations.

\subsection{\texorpdfstring{Large $N$ evaluation of the SCI}{}}
\label{largeN}

In this section we compute the large $N$ behavior of the new descendant solutions. This can be achieved by employing the standard techniques developed in \cite{Benini:2018ywd, Lanir:2019abx, Benini:2020gjh, Aharony:2021zkr}. Here, we recall the main aspects necessary for the evaluation.

We start by rewriting the index \eqref{SCIneck} using the gauge holonomies $u_i^{(a)}$\footnote{We define conventionally $u^{(k+1)}=u^{(1)}$.} and the flavor fugacities  $\Delta_{a,b}$:
\begin{equation}
	\label{SCI2}
	\begin{aligned}
		\mathcal{I} = 
		\left(\frac{(q;q)_{\infty}^{2(N-1)} }{N!}\right)^{\!K}
		\prod_{a=1}^K \widetilde \Gamma_e\left( \Delta_{a,a};\tau \right)^{(N-1)}
		\oint_{\mathbb{T}^{K(N-1)}}\prod_{a=1}^K
		\left(\,
		 \prod_{i=1}^{N-1}
		\dd{u_i^{(a)}} 
		 \mathcal{Z}_a(u,\Delta;\tau)\right) ,
	\end{aligned}
\end{equation}
where we have introduced the modified elliptic gamma function $\widetilde{\Gamma}_e(x) = \Gamma_e(\ee^{2\pi \mi x})$ and the integral runs over the gauge holonomies $u^{(a)}_i$, collectively denoted as $u$. The integrand $\mathcal{Z}_a$ is defined by
\begin{equation}
		\mathcal{Z}_a(u,\Delta;\tau)=\prod_{i \ne j}^{N}
		\frac{
			\widetilde{\Gamma}_e\left(\Delta_{a,a} +u_{ij}^{(a,a)}\right)
		}{\widetilde{\Gamma}_e\left(u_{ij}^{(a,a)}\right)}
		\prod_{i, j=1}^{N}
		\!\widetilde{\Gamma}_e\left(\Delta_{a,a + 1} +u_{ij}^{(a,a+1)}\right)
		\!\widetilde{\Gamma}_e\left(\Delta_{a + 1, a}+ u_{ij}^{(a + 1,a)}
		\right)\!,
\end{equation}
where we used the shorthand notation \eqref{notation} for $u_{ij}^{(a,b)}$. The $\Delta_{a,b}$ parameters are constrained by the superpotential at each node of the quiver as in \eqref{Dconstr} with $\tau=\sigma$.

\begin{comment}
\begin{align}
    \label{deltaconst}
    \Delta_{a,a} + \Delta_{a,a+1} + \Delta_{a+1,a} = 2\tau, \qquad \forall a=1,..., K, \nonumber \\
    \Delta_{a,a} + \Delta_{a,a-1} + \Delta_{a-1,a} = 2\tau, \qquad \forall a=1,..., K.
\end{align}
\end{comment}

We aim to compute the contribution to the index of our set of descendant solutions to the BAEs, introduced in Section \ref{NewSol}. Consider the descendant $\{m,n,r\}_s$ \eqref{HLgen}, we highlight that the dependence on the quiver structure of these solutions is fully incorporated in $\mathbf{s}^{(a)}$, while information on the gauge group is encoded in the remaining terms of \eqref{HLgen}. As a consequence, computing the difference
\begin{equation}
   \hat{u}_{ij}^{(a,b)}\equiv \hat{u}_i^{(a)}-\hat{u}_j^{(b)} = \hat{u}^{\mathrm{HL}}_{ij} + \delta^{(a,b)},
\end{equation}
we can isolate a node-independent term $\hat{u}^{\mathrm{HL}}_{ij}$, obtained by evaluating the quantity on the HL progenitor, and therefore common to the entire family, as well as a part $\delta^{(a,b)} ={s}^{(a)}-{s}^{(b)}$, which encodes the dependence on the descendant solution. Such term measures the relative difference in $\bar{u}^{(a)}$ between two gauge nodes and as such only contributes to the bifundamental hypermultiplets as
\begin{equation}
    \widetilde{\Gamma}_e\left(\Delta_{a,b} + \delta^{(a,b)} + \hat{u}_{ij}^{\mathrm{HL}}\right).
\end{equation}

More specifically, in the case at hand, these terms are non-vanishing only for matter fields connecting consecutive nodes and, in the effective theory obtained by compactification on the circle $S^1_\tau$ of size $\tau$, act as large vector-like baryonic real-mass parameters for the corresponding hypermultiplets. These masses are associated with the (decoupled) $\mathrm{U}(1)$ factors at the various nodes, upon interpreting the index as the supersymmetric partition function of the theory on $S^1_\tau\times S^3$. This is also consistent with the brane picture of the model, described in Section \ref{necklacereview}, where the relative distance between each pair of consecutive D4-branes acts as a bare mass parameter for the corresponding hypermultiplet.

Then, it  is natural to introduce generalized  $\hat{\Delta}_{a,b}$ parameters 
\begin{equation}
 \hat{\Delta}_{a,b} = \Delta_{a,b} + \delta^{(a,b)}.
\end{equation}
Notice that $\hat{\Delta}_{a,a} = \Delta_{a,a}$ and $\hat{\Delta}_{a,a+1} +\hat{\Delta}_{a+1,a} = \Delta_{a,a+1} +\Delta_{a+1,a}$, consistent with it being a baryonic real mass. Consequently, the $\hat\Delta$ satisfy the same constraint as the $\Delta$.
The contribution to $\mathcal{Z}_a$ of a solution \eqref{HLgen} takes the form
\begin{equation}
\label{nodehlcont}
		\mathcal{Z}_a(u,\Delta;\tau)=\prod_{i \ne j}^{N}
		\frac{
			\widetilde{\Gamma}_e\left(\hat{\Delta}_{a,a} + \hat{u}_{ij}^{\mathrm{HL}}\right)}{\widetilde{\Gamma}_e\left(\hat{u}_{ij}^{\mathrm{HL}}\right)}
		\prod_{i, j=1}^{N}
		\widetilde{\Gamma}_e\left(\hat{\Delta}_{a,a+1} + \hat{u}_{ij}^{\mathrm{HL}}\right)
		\widetilde{\Gamma}_e\left(\hat{\Delta}_{a+1,a} + \hat{u}_{ij}^{\mathrm{HL}}
		\right),
\end{equation}
with
$\hat{\Delta}_{a,a}+\hat{\Delta}_{a,a\pm1}+\hat{\Delta}_{a\pm1,a}=2\tau$, \emph{i.e.} they still satisfy the same constraint \eqref{Dconstr} of the non-hatted parameters. 

We observe that the shift $s$ only enters as difference between distinct nodes. This is consistent with the fact that only relative differences are associated with  real masses of hypermultiplet, while global shifts, associated with the diagonal $\mathrm{U}(1)$,  as discussed in the previous sections, carry unphysical information, resulting in an overall degeneracy factor $N$ for the contribution of each solution to the index. 

The term \eqref{nodehlcont}, associated to each node $a$, has now the same form of the integrand of $\mathrm{SU}(N)$ $\mathcal{N} = 4$ SYM evaluated on a standard HL progenitor $\{m,n,r\}$ with charges for the matter fields
\begin{equation}
    \hat{\Delta}_1\equiv \hat{\Delta}_{a,a+1}, \quad \hat{\Delta}_2\equiv \hat{\Delta}_{a+1,a},\quad \hat{\Delta}_3\equiv \hat{\Delta}_{a,a} =2\tau - \hat{\Delta}_1 - \hat{\Delta}_2, 
\end{equation}
upon also including the zero modes $\widetilde{\Gamma}(\hat{\Delta}_{a,a})^{N-1}$ for the adjoint matter fields at each node,
plus two extra zero modes arising from the bifundamental fields
\begin{equation}
\label{extrazero}
    \mathcal{Z}_a(\hat{u},\hat{\Delta}) =  \mathcal{Z}^{\mathcal{N}=4}_{a} (\hat{u},\hat{\Delta})    \,\widetilde\Gamma(\hat{\Delta}_{1})
    \,\widetilde\Gamma(\hat{\Delta}_{2})\,. 
\end{equation}
The term $\mathcal{Z}_a^{\mathcal{N}=4}$ has already been evaluated on the HL solutions $\{m,n,r\}$ in \cite{Aharony:2021zkr} and amounts to 
\begin{align}
\label{sunsym}
   \log \biggl(\frac{(q;q)_\infty^{2(N-1)}}{N!}\mathcal{Z}^{\mathcal{N} = 4}_a&\left(\{m,n,r\},\hat{\Delta}\right)\biggr) \!=\! \nonumber \\
   &=
   -\frac{\pi \mi N^2}{m} \frac{([m \hat{\Delta}_1]_{\check{\tau}}+n)([m \hat{\Delta}_2]_{\check{\tau}}+n)([m \hat{\Delta}_3]_{\check{\tau}}+n)}{\check{\tau}^2} +\mathcal{O}(N)\,,   
\end{align}
with the constraint
\begin{equation}
\label{Vincolo}
\sum_{a = 1}^{3}[m\hat{\Delta}_a]_{\check{\tau}} = 2\check{\tau} - 1-n\,,  \qquad n=0,1\,,
\end{equation}
where $\check{\tau} = m \tau + r$.
We refer the reader to \cite{Benini:2018ywd, Aharony:2021zkr} for the explicit computation, but highlight two fundamental aspects in the evaluation, which are necessary to understand the salient features of the resulting expression and the definition of $[\, \cdot \,]_{\check{\tau}}$.
\begin{itemize}
    \item  On a HL solution employing the periodicity of the elliptic gamma functions, each $\widetilde{\Gamma}_e(\hat{\Delta} + u_{ij})$ takes the form

\begin{equation}
   \prod_{i\neq j}\widetilde{\Gamma}_e\left(\hat{u}_{ij}^{\mathrm{HL}}+ \hat{\Delta}\right) =  \prod_{I=0}^{m-1}\prod_{J,J_0=0}^{n-1}\widetilde{\Gamma}_e\left(\frac{I}{m}+ \frac{J-J_0}{N} \check{\tau} + \hat{\Delta} \right)^m,
\end{equation}
with $I=0,...,m-1$ and $J,J_0 =0,...,n-1$. The product over $I$ can now be evaluated exactly yielding
\begin{equation}
    \prod_{I=0}^{m-1}\prod_{J,J_0=0}^{n-1}\widetilde{\Gamma}_e\left(\frac{I}{m}+ \frac{J-J_0}{N} \check{\tau} + \hat{\Delta}; \tau \right)^m = \prod_{J,J_0=0}^{n-1}\widetilde{\Gamma}_e\left(\frac{J-J_0}{n} \check{\tau} + m\hat{\Delta}; \check{\tau} \right)^m.
\end{equation}
Such evaluation changes the modular parameter $\tau$ into $\check{\tau}$ naturally associated with the corresponding HL solution. 
\item The elliptic gamma function with $\tau = \sigma$ admits the infinite product expansion \cite{Felder:1999qhl}
\begin{equation}
\label{gammainf}
\widetilde{\Gamma}_e(x;\tau) = \frac{\ee^{-\pi\mi Q(x,\tau)}}{\theta_0\left(\frac{x}{\tau};-\frac{1}{\tau})\right)}\prod_{n = 0}^{\infty}\frac{\psi\left(\frac{n+x+1}{\tau}\right)}{\psi\left(\frac{n-x}{\tau}\right)}\,,
\end{equation}
with $\theta_0(z;q) = (z;q)_{\infty} (q/z;q)_{\infty}$, while 
\begin{equation}
\label{specialpeto}
\begin{aligned}
    \psi&(t) = \exp \left(
    t \log \left(1 -  \ee^{-2 \pi \mi t}\right) - \frac{1}{2 \pi \mi} \mathrm{Li}_2 \left(\ee^{2 \pi \mi t} \right)
    \right) \,, 
    \qquad 
    \mathrm{Li}_2 (z) = \sum_{n=1}^\infty \frac{z^n}{n^2}\,, \\
    & 
    Q(u;\tau) = 
    \frac{1}{3 \tau^2} u^3-
    \frac{2 \tau - 1}{2 \tau^2} u^2 +
    \frac{5 \tau^2 - 6 \tau + 1}{6 \tau^2} u - 
    \frac{(2 \tau -1)(\tau-1) }{12 \tau}\,.
\end{aligned}
\end{equation}
On a $\{m,n,r\}$ solution in the large $N$ limit, with $m,r$ fixed and $n\to\infty$ this expansion leads to the following estimate 
\begin{equation}
    \sum_{i\neq j}\log \widetilde{\Gamma}_e\left(\hat{u}_{ij}^{\mathrm{HL}}+\hat{\Delta};\tau\right) \sim - \frac{\pi \mi  N^2}{m}\left(Q(m \hat{\Delta},\check{\tau}) + 12 \check{\tau}^2\, Q''(m \hat{\Delta},\check{\tau})\right) + \mathcal{O}(N),
\end{equation}
which holds in the region $ 0 < \Im \frac{m \hat\Delta}{\check{\tau}} < -\Im \frac{1}{\check{\tau}}$.
Since the elliptic gamma function $\widetilde{\Gamma}_e(x;\tau)$ is actually periodic under $x \to x+1$, the result can be extended to all $\mathbb{C} \setminus \mathcal{L} $, where $\mathcal{L}$ is the set of lines bounding the domain where the previous estimate holds. To achieve this, one defines $[m \Delta]_{\check{\tau}}$ as the representative in $ 0 < \Im \frac{m \hat\Delta}{\check{\tau}} < -\Im \frac{1}{\check{\tau}}$ of the equivalence class $m\hat{\Delta} \sim m\hat{\Delta} + n$, with $n \in \mathbb{Z}$. After summing over $\Delta_{a = 1,2,3}$ and employing the constraint imposed by the superpotential, one finally recovers the cubic expression \eqref{sunsym}. Notice that because of the superpotential constraint \eqref{Vincolo} we have
\begin{equation}
    [m\hat{\Delta}_3]=2\tau -1 -[ m\hat{\Delta}_1 + m\hat{\Delta}_2]_{\check{\tau}}.
\end{equation}
Depending on wether 
\begin{equation}
    [m\hat{\Delta}_1 + m\hat{\Delta}_2]_{\check{\tau}} = [m \hat{\Delta}_1]_{\check{\tau}}+[m \hat{\Delta}_2]_{\check{\tau}}+n, \quad \text{with} \quad n = 0, 1.
\end{equation}
there are two cases.
\end{itemize}

In order to understand the fate of the extra two zero modes, we recall that inside $\mathcal{O}(N\log N)$ there is a $\mathcal{O}(1)$ term of the form
\begin{equation}
\label{symu1}
    -\log \mathcal{I}_{\mathrm{U}(1)} = -\log\left((q;q)^2_\infty\,\, \widetilde{\Gamma}_e(\hat{\Delta}_1) \,\widetilde{\Gamma}_e(\hat{\Delta}_2) \,\widetilde{\Gamma}_e(\hat{\Delta}_3)\right),
\end{equation}
associated to the index of a free $\mathcal{N} = 4$ $\mathrm{U}(1)$ SYM as discussed in \cite{Aharony:2021zkr}. Combining \eqref{extrazero} and \eqref{symu1}, the two zero modes cancel out leaving only a single zero mode in the denominator of $\mathcal{I}$, associated with the decoupled $\mathrm{U}(1)$ factor for the single adjoint that is left on each node of the model after the orbifold. 

Finally, we recall that each solution brings a degeneracy factor $N \cdot N!^K$. The $N!^K$ cancels precisely with the Weyl group factor leaving an overall $N$ degeneracy factor associated with the diagonal $\mathbb{Z}_N$ factor as discussed in the previous sections. 

We now need to combine together the contributions of each node. 
Using the superpotential constraints for each node $a$, we can express the adjoint $[m\hat{\Delta}_{a,a}]_{\check{\tau}}$ in terms of the bifundamentals as
\begin{equation}
    [m\hat{\Delta}_{a,a}]_{\check{\tau}} = 2 \check{\tau} - 1 - n_{(a)} -[m\hat{\Delta}_{a,a+1}]_{\check{\tau}} -[m\hat{\Delta}_{a+1,a}]_{\check{\tau}}, \qquad \forall a=1,...,K
\end{equation}
with each $n_{(a)}=0,1$.
Summing the contributions $\mathcal{Z}_a$ from each node we obtain
\begin{equation}
\log \mathcal{Z} = - \frac{\pi \mi N^2}{m \check{\tau}^2} \sum_{a=1}^K \left([m \hat{\Delta}_{a,a}]_{\check{\tau}}+n\right)\left([m \hat{\Delta}_{a,a+1}]_{\check{\tau}}+n\right)\left([m \hat{\Delta}_{a+1,a}]_{\check{\tau}}+n\right)  +\mathcal{O}(N)\,.
\end{equation}
To complete the evaluation of the index using \eqref{baf} we would still need to evaluate the Jacobian $\mathcal{H}$. However, employing standard arguments discussed in \cite{Benini:2018ywd, Aharony:2021zkr}, such a term only contributes at subleading order $\mathcal{O}(N\log N)$ in the large $N$ limit. Therefore, at leading order the full index evaluates on the descendant solutions to
\begin{equation}
\label{indexlargen}
\begin{aligned}
\log \mathcal{I} = \!- \frac{\pi \mi N^2}{m \check{\tau}^2}\! \sum_{a=1}^K\! \left([m \hat{\Delta}_{a,a}]_{\check{\tau}}+n_{(a)}\right)\!\! \left([m \hat{\Delta}_{a,a+1}]_{\check{\tau}} + n_{(a)}\right)\!\! \left([m \hat{\Delta}_{a+1,a}]_{\check{\tau}}\right. &+ n_{(a)}\Bigr) \\
&+ \,\mathcal{O}(N\log N)\,,
\end{aligned}
\end{equation}

We could have obtained this very same result also by noticing that, because HL progenitors are independent on the parameters $\Delta$ we can choose to study the index (and the BAEs) at the point in the parameter space $\hat{\Delta}$ fixed as above. Then, since $\delta^{(a,b)}=s^{(a)}-s^{(b)}$ does not depend on the gauge indices we can define $u_i^{(a)} = u_{i}^{\mathrm{HL}} + s^{(a)}$ and obtain the following relation
\begin{equation}
    I\!\left(\hat{u}^{\mathrm{HL}},\hat{\Delta};\tau\right) = I\left(\hat{u}(m,n,r,s),\Delta;\tau\right)\,,
\end{equation}
which shows how any descendant \eqref{HLgen} can be seen as an HL progenitor on a shifted (baryonic) background in the $\Delta$ parameters. Using this result, one immediately recovers \eqref{indexlargen} from the HL expression. Moreover, applying the same argument to the BAEs \eqref{NKbaesTH1} evaluated on the HL solutions provides an alternative derivation of the solutions \eqref{HLgen}, with the lattice condition on \(s^{(a)}\) arising directly from the \(\mathrm{SU}(N)\) constraint.

Out of the $3K$ parameters, only $K+1$ are independent, since the $\hat{\Delta}_{a,b}$ provide a parameterization for the global symmetries of the theory. 
Starting from \eqref{Dconstr} and combining the equations for the node $a$ with the ones for the node $a+1$ we obtain 
\begin{equation}
\label{adj}
\hat{\Delta}_{1,1} = \hat{\Delta}_{2,2}= ... = \hat{\Delta}.
\end{equation}
Only $K-1$ of these equations are independent. Together with the $K$ equations
\begin{equation}
\label{constradj}
       {\Delta}_{a,a} = 2 \tau -\hat{\Delta}_{a,a+1} -\hat{\Delta}_{a+1,a}\,, \qquad \forall\; a=1,...,K
\end{equation}
they imply a corresponding constraint on the bifundamental fields
\begin{equation}
\label{constrbif}
    \hat{\Delta}_{a,a+1}+ \hat{\Delta}_{a+1,a} = \hat{\Delta}_{a-1,a}+ \hat{\Delta}_{a,a-1}\,, \qquad \forall \; a=1,...,\, K-1.
\end{equation}
All in all, \eqref{adj} and \eqref{constradj} are $2K+1$ independent constraints on the parameters $\hat{\Delta}_{a,b}$.

When passing to the $[m\hat{\Delta}_{a,b}]_{\check{\tau}}$ the linearly dependent constraints \eqref{constrbif}, become now consistency conditions which our parameters must satisfy, dividing the parameters space in $2^K$ regions, labeled by the possible choices of $n_{(a)}$ on each node.
Summarizing, we get
\begin{align}
    &[m\hat{\Delta}_{a,a}]_{\check{\tau}} = 2 \check{\tau} - 1 - n_{(a)} -[m\hat{\Delta}_{a,a+1}]_{\check{\tau}} -[m\hat{\Delta}_{a+1,a}]_{\check{\tau}}\,, \qquad && \forall \;a=1,...,K\,, \nonumber \\
    &[m\hat{\Delta}_{a,a}]_{\check{\tau}} = [m \hat{\Delta}_{a-1,a-1}]_{\check{\tau}}\,, \quad && \forall \; a=1,...,K-1\,,
\end{align}
together with the consistency conditions
\begin{equation}
\label{consistency}
    [m\hat{\Delta}_{a,a+1}] + [m\hat{\Delta}_{a+1,a}] + n_{(a)}-n_{(a-1)} = [m\hat{\Delta}_{a-1,a}]+ [m\hat{\Delta}_{a,a-1}]\,.
\end{equation}

We now notice that despite having $2^K$ regions, many of them produce an equivalent physical description upon a renaming of the chemical potentials, \textit{i.e.} acting with the symmetry group of the necklace quivers, the affine Weyl group $\hat{A}_{K-1}$. For instance, we can classify the regions with a vector $(n_{(1)}, n_{(2)},...,n_{(K)})$. The regions labeled by vectors related by permutations describe an equivalent physical situation. Therefore, modulo ordering we have precisely $K+1$ distinct regions. This is consistent with the results of \cite{Benini:2020gjh} for which the number of regions in the parameter space of the superconformal index of any $\mathcal{N} = 1$ toric model is given by the number of vectors of the toric fan, related to the number of global symmetries. For $\mathbb{C}^2/\Z_K\times \mathbb{C}$ there are precisely $K+1$ vectors, including integer points on the edges of the toric diagram.
The leading order $N^2$ term is consistent with the expected scaling behavior of toric models \cite{Benini:2020gjh}, and captures holographically the entropy of large black holes in the dual gravitational setup. 
Notice that our formula reduces to the known formula for the HL solutions for $\mathcal{N} = 4$ SYM, corresponding to $K=1$, as well as the general formula valid for any toric quiver gauge theory \cite{Benini:2020gjh} for the basic HL solution $\{1,N,0\}$, for which $\hat{\Delta}_{a,b} = \Delta_{a,b}$.
In the region $(0,0,...,0)$, \eqref{indexlargen} takes the particularly simple form
\begin{equation}
   \log \mathcal{I} = -\frac{ \pi \mi N^2 }{m \check{\tau}^2} [m \hat{\Delta}]_{\check{\tau}}\sum_{a=1}^K\,  [m \hat{\Delta}_{a,a+1}]_{\check{\tau}}  [m \hat{\Delta}_{a+1,a}]_{\check{\tau}}\,.
\end{equation}

\section{Conclusions}
\label{secconc}
In this work we studied the application of the Bethe Ansatz approach to the case of $\mathcal{N}=2$ necklace theories with $K$ $\SU(N)$ gauge nodes.
Our findings generalize what was already known for $\mathcal{N}=4$ $\SU(N)$ SYM and for non chiral toric theories.
Indeed, we proposed an ansatz to build new families of descendant discrete solutions for the BAEs of the necklace theories starting from the progenitor HL solutions and then we proved that the descendant solutions solve the BAEs.
Moreover our ansatz extends beyond the necklace theories, as the new Bethe vacua solve the BAEs for any non chiral toric theory with $\SU(N)$ gauge nodes.

We explored the features of these new families of Bethe vacua.
The combined number of inequivalent descendant and progenitor solutions  coincides with the number of massive vacua of the $\mathcal{N}=1^*$ massive deformations of the necklace theory.
Such correspondence is in agreement with the conjecture of \cite{ArabiArdehali:2019orz,Benini:2021ano}, originally formulated for $\mathcal{N}=4$ $\SU(N)$ SYM.

Moreover, we studied the action of the S-duality group on the discrete solutions of the BAEs, where the S-duality group takes the form of the mapping class group $\mathcal{M}(1,K)$ of the $K$-punctured torus.
In the case of $\mathcal{N}=4$ $\SU(N)$ SYM (corresponding to the case $K=1$ here), the  solutions sit into orbits of the S-duality group $\mathrm{SL}(2,\mathbb{Z})$.
We showed that such structure also holds for the necklace theories with $\SU(N)^K$ (with $K>1$), when considering all the inequivalent descendant and progenitor solutions.

We also discussed the evaluation of the contributions to the SCI coming from these new descendant solutions, both for finite and large $N$.
In the finite $N$ regime, we showed a concrete example where such contributions are still not sufficient to fully reproduce the SCI.
Differently from the case of the $\SU(2)^2$ conifold theory and the $\SU(2)^3$ SPP theory studied in \cite{Amariti:2025vjd}, the computation of the SCI of $\SU(2)^2$ necklace theory requires the contributions of the continuous solutions.
This suggests  that the contributions of continuous solutions become irrelevant when we turn on suitable masses deformations to the adjoint chirals.
Finally, we computed the large $N$ limit of the contribution of the new families of solutions for an arbitrary number of gauge nodes.

It would be interesting to further investigate the correspondence between the Bethe problem and the associated integrable system \cite{Dorey:2001qj,Hollowood:2002zk}.
Our findings strengthen the correspondence between the numbers of solutions of the two problems.
A more explicit understanding could be obtained by constructing a map between the two, which would allow results obtained in one framework to be translated into the other. 
For example such a relation has been recently discussed in \cite{Fazzi:2026wkb},
where a map between the BAEs of $\SU(N)$ $\mathcal{N}=4$ SYM and the 
extremization equations of the $A_{N-1}$ elliptic Calogero-Moser system was provided.
In our case a similar analysis is actually more intricate, because the  equations are  more complicated and  there are extra variables that do not have a clear interpretation from the BAEs perspective. In such cases indeed the vacua have been found through an educated ansatz in \cite{Dorey:2001qj} for general $N$ and $K$.
The situation simplifies in the $\SU(2)\times \SU(2)$ case, that was analyzed in full generality in \cite{Hollowood:2002zk}.
In such case, one should identify the relevant extremization equations for the elliptic spin Calogero-Moser system that can be put in correspondence of the Bethe problem. 
Restricting to the discrete solutions  we have found promising preliminary results.
On the other hand we think that the analysis of the extremization equation of the elliptic spin Calogero-Moser system can shed some light on the structure of the continuous branch as well.
We aim to report on results in this direction in the next future.

\section*{Acknowledgments}
We thank E. Colombo, M. de Marco, R. Argurio, K. Krawczyk and M. Fazzi for useful discussions. This work of A.A., P.G. and C.M. has been supported in part by the Italian Ministero dell'Istruzione, Università e Ricerca (MIUR), in part by Istituto Nazionale di Fisica Nucleare (INFN) through the “Gauge Theories, Strings, Supergravity” (GSS) research project. The work of A.Z. has been supported by ``Fondazione Angelo Della Riccia”.

\appendix

\section{\texorpdfstring{$\mathcal{N}=1^*$ SYM}{}}
\label{AppN=1*}
In this Appendix we summarize aspects about $\mathcal{N} = 1^*$ SYM the classification of vacua of the theory.

$\mathcal{N} = 1^*$ SYM is obtained from a massive $\mathcal{N} = 1$-preserving deformation of $\mathcal{N} = 4$ SYM, by adding the following mass terms for the chiral adjoint fields $X_{1,2,3}$ to the superpotential
\begin{equation}
	W = \Tr X_1 \left[X_2, X_3\right] + \frac{1}{2} \sum_{a=1}^3 m_a X_a^2\,.
\end{equation}
The vacua structure of this theory was studied in \cite{Donagi:1995cf,Bourget:2016yhy, Bourget:2015lua} by studying $F$- and $D$-terms equation, which for $\mathcal{N}=1$ theories is equivalent to the study of $F$-term modulo complexified gauge transformations $\mathrm{SL}(N,\mathbb{C})$. The equations define a $\mathfrak{su}(2)$ algebra for $X_a$ $a=1,...,3.$ upon modding out $\mathrm{SL}(N,\mathbb{C})$. The vacua are then classified by the conjugacy classes of the homomorphism $\rho:\mathfrak{su}(2)\to \mathfrak{su}(N)$, which are in one-to-one correspondence with the nilpotent orbits of $\mathfrak{g}=A_{N-1}$. In turn, the orbits of $A_{N-1}$ are in biijections with the partitions of $N$. Each partition is then associated with some $\rho$, which will Higgs the original gauge group down to a subgroup $H$. The vacua and phases of the theory can be organized depending on the nature of $H$. 
\begin{itemize}
\item For equipartitions $N = m n$ the vacua are massive. For each divisor $d$ of $N$, $H = \mathrm{SU}(d)$, the low-energy theory is pure $\mathcal{N}=1$ SYM with gauge group $H$ as the matter field have bare masses and will be integrated out. The theory exhibits confinement and has precisely $m$ vacua. The extremal partition $N = 1\cdot N$ is a Higgs vacuum, with $G$ completely broken. The total number of these vacua is given by $\sigma_1(N)$. Interestingly, such vacua sit in orbits of $\mathrm{SL}(2,\mathbb{Z})$ as it happens for the HL solutions. In this case this is due to the $\mathrm{SL}(2,\mathbb{Z})$ S-duality group of the original $\mathcal{N}=4$ SYM, which acts in the $\mathcal{N} = 1^*$ deformation, by permuting these vacua among themselves. As an immediate consequence, in this theory there is a one-to-one correspondence between the number of massive vacua, global structures, phases of the theory and lattices of mutually probe line operators which detect such phases \cite{Aharony:2013hda}.
\item Generic partitions leave abelian factors of the gauge group $G$. The number $l$ of photons is given by the number of such abelian factors. There is no mass gap for these vacua so they are massless branches of vacua describing a Coulomb phase of the theory with $\ell$ photons.
\end{itemize}

\bibliographystyle{JHEP}
\bibliography{refHL.bib}
\end{document}